\documentclass[12pt]{article}
\usepackage{graphicx,bm,amsmath,amssymb,latexsym,psfrag,epsfig,subfigure,euscript, multirow,color,verbatim,bbm,cancel}

\usepackage[margin=2cm]{geometry}
\usepackage{appendix}

\allowdisplaybreaks
\newcommand{\MPl}{\overline{M}_{\mathrm{Pl}}}
\newcommand{\Mg}{M_{\mathrm{g}}}
\newcommand{\Mzp}{M_{Z'}}

\newcommand{\nn}{\nonumber \\}
\newcommand{\ds}{\displaystyle }
\newcommand{\e}{\varepsilon }

\begin{document}
\begin{titlepage}

\begin{flushright}
\end{flushright}

\vspace{0.2cm}
\begin{center}
\Large\bf
Early (and Later) LHC Search Strategies for Broad Dimuon Resonances 
\end{center}

\vspace{0.2cm}
\begin{center}
{\sc Randall Kelley, Lisa Randall, and Brian Shuve}\\
\vspace{0.4cm}
{\sl Department of Physics\\
Harvard University\\
Cambridge, MA 02138, U.S.A.}
\end{center}

\vspace{0.2cm}
\begin{abstract}\vspace{0.2cm}
\noindent 
Resonance searches generally focus on narrow states that would produce a sharp peak rising over background. Early LHC running will, however, be sensitive primarily to broad resonances. In this paper we demonstrate that statistical methods should suffice to find broad resonances and distinguish them from both background and contact interactions over a large range of previously unexplored parameter space. We furthermore introduce an angular measure we call ellipticity, which measures how forward (or backward) the muon is in $\eta$, and
allows for discrimination between
models with different parity violation early in the LHC running.  We contrast this with existing angular observables
and demonstrate that ellipticity is superior for discrimination based on parity violation, while others
are better at spin determination.

\end{abstract}
\vfil

\end{titlepage}


\section{Introduction}
With the LHC running smoothly and the ``re-discovery'' of the Standard Model (SM) well underway,  
it is time to contemplate what new physics this early LHC run might access.    A significant 
amount of data ($\sim 1\,\,\mathrm{fb}^{-1}$) will be accumulated at 
$\sqrt s = 7\,\,\mathrm{TeV}$, and it is important to consider what kinds of physics beyond the SM 
can be studied at this energy and luminosity, particularly with several years' wait until 
the anticipated $14\,\,\mathrm{TeV}$ run.   The models of greatest interest in this regard are those that will be accessible to  early  LHC running but are outside of the range of  the Tevatron.

The first and simplest target for new physics searches will be resonance searches. The
Drell-Yan channel into leptons will likely be the simplest channel to study, since the SM contributions are well-understood,
the background is very low at high invariant mass, and many models of new physics have sizeable
branching fractions to leptons, particularly muons.
Our focus will therefore be on muons, though we  will also briefly consider the additional
contributions from electrons or photons that could be present and  increase  discovery reach.  
Muon detector efficiencies and resolutions are very good at both CMS and ATLAS, as are those for electrons.   
However, models involving strong electron couplings are generally tightly constrained by LEP,   
making dimuon searches more interesting in those cases.
The production of
strongly-interacting final states (such as $t\bar t$) could also be of interest, as they are  
enhanced by a color factor, but
this advantage is typically negated by higher QCD backgrounds,
lower efficiencies, and worse resolution, so
we will not consider these further.

The $p\bar p$ nature of the Tevatron ensures that it will serve as the best probe of neutral 
resonances below 1 TeV.  The LHC will nonetheless dominate for other cases due to
the higher center-of-mass (CM)
energy. This makes high-mass resonances, which were out of the Tevatron's reach, kinematically accessible. Furthermore, the large gluonic parton fraction 
at low $x$ means that resonances coupling to gluons (such as Kaluza-Klein
 (KK) gravitons in the Randall-Sundrum (RS) framework)
 will have  enhanced cross sections over the Tevatron.    Even so, to be visible with only $1\,\,\mathrm{fb}^{-1}$ of data,
 large couplings to leptonic final states are most likely essential.  Since the width of a 
resonance scales as the square of the coupling, this implies that, during the early LHC running,
we will find broad resonances with large decay widths.

Broad resonances are more challenging to study than narrow ones.  With narrow resonances, 
the invariant mass of the signal events are tightly clustered around the particle's mass. 
For muons, the detector resolution is sufficiently good that performing a simple counting experiment 
in the region around the resonance pole (with a width on the order of the muon resolution) often gives a signal-to-background ratio 
that is high enough for detection.  Detecting broad resonances will be more challenging since
signal events are 
spread out over a wider area so the signal will not be as clearly identifiable
by its shape.  Distinguishing signal from background is further complicated by parton
distribution uncertainties in the high 
invariant mass region, as well as uncertainties in the integrated luminosity, both of which are often
estimated using the Drell-Yan process that we're using as a probe of new physics.

The question then
becomes whether or not we can distinguish these events from background or other forms of
strongly-interacting physics such as contact interactions. In this paper we show that characteristic features 
of the invariant mass distribution will be sufficient to find and identify resonances, even at the first run of the LHC.
   In its simplest form, our method involves looking for an absolute 
rise in the differential cross section, or ``upturn'', which would clearly distinguish any new physics
from the SM or a contact interaction, both of which predict a falling distribution.
Our fuller,
more sophisticated statistical analysis uses the maximum likelihood method to distinguish between
resonances and contact interactions (of which the SM is the limiting case where we take the contact
interaction scale $\Lambda\rightarrow\infty$) and covers a greater parameter range.  We demonstrate
that differentiating a broad resonance from background
and contact interactions is possible over much of the accessible range of parameter space with a significant number of signal events.

We also use angular distributions to further distinguish new physics models
from one another and from the SM.
We study the muon pseudorapidity ($\eta^-$)  distribution,
and demonstrate that a variable we call
ellipticity can be used to distinguish among different chiral structures.
We find that ellipticity is good at distinguishing models
with different parity violation, whereas observables based on
the muon angle in the CM frame ($\theta^*$)
are more useful for spin determination.
Model discrimination based on angular information generally requires more 
events than will be accessible in the first LHC run for
resonance masses larger than 1 TeV, so we also consider
the prospects of model discrimination with $\sqrt s = 10\,\,\mathrm{TeV}$
and different integrated luminosities.

Our analysis follows others who have investigated the LHC reach of $Z'$ gauge
bosons during the $\sqrt s = 7\,\,\mathrm{TeV}$ run with various integrated
luminosities \cite{Bauer:2009cc,Diener:2009vq,
Salvioni:2009mt,Salvioni:2009jp,Basso:2010pe}.  We extend these analyses to
include searches for RS KK gravitons, and widen the study to include
  strongly-coupled, broad resonances.


\section{Resonance or Contact Interaction?} \label{sec:resorcontact}
We begin our work by motivating the fundamental question addressed in this paper:
can we distinguish resonances from contact interactions in the 7 TeV
run of the LHC?  For concreteness, we examine $Z'$ models and KK gravitons within
the RS framework (we will describe the details of these models in 
section \ref{sec:models}).  As we will demonstrate in section \ref{sec:xsec},
the largest accessible regions of parameter space in the early running of the
LHC will be strongly coupled, resulting in a broad resonance.

\begin{figure}[t]
\begin{center}
\includegraphics[scale=0.8]{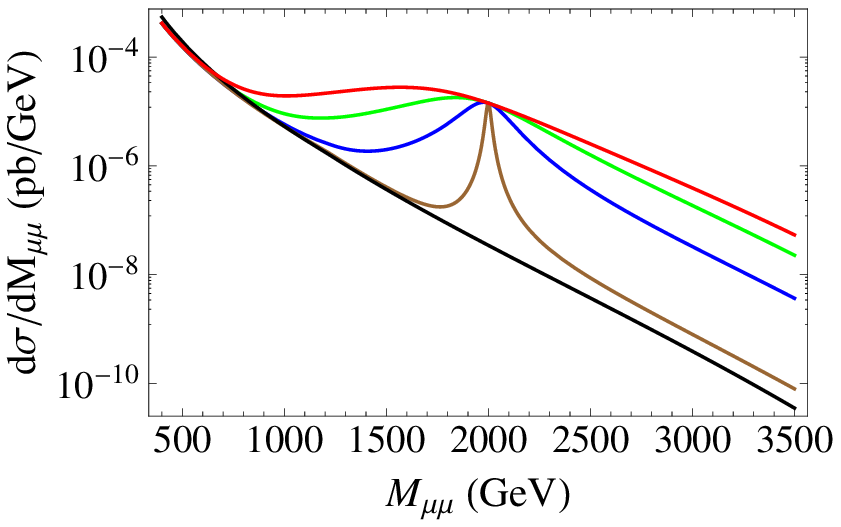}
\includegraphics[scale=0.8]{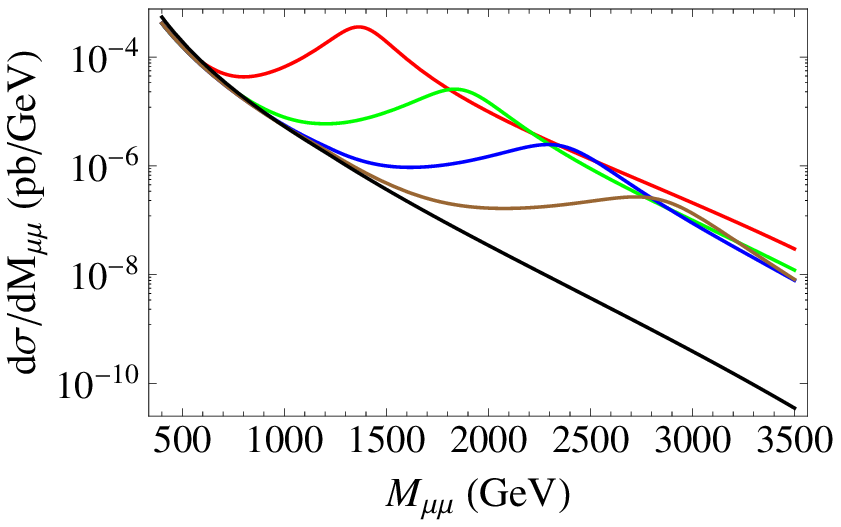}\\
\includegraphics[scale=0.8]{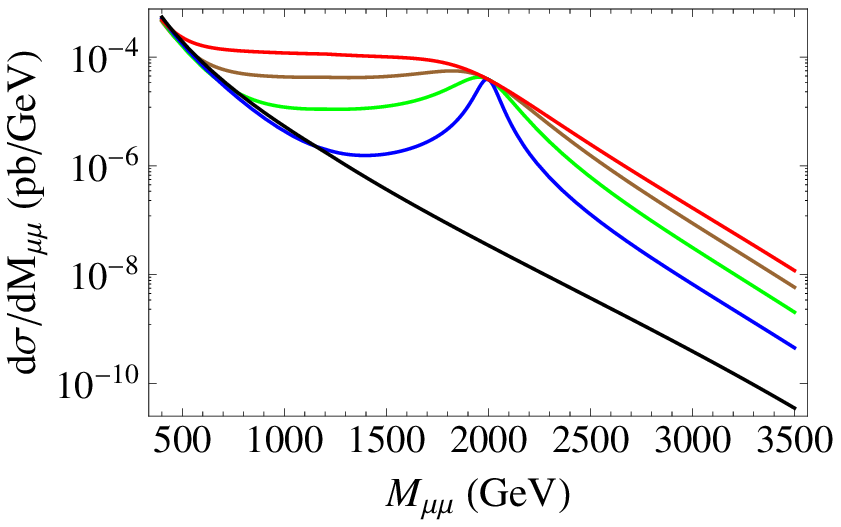}
\includegraphics[scale=0.8]{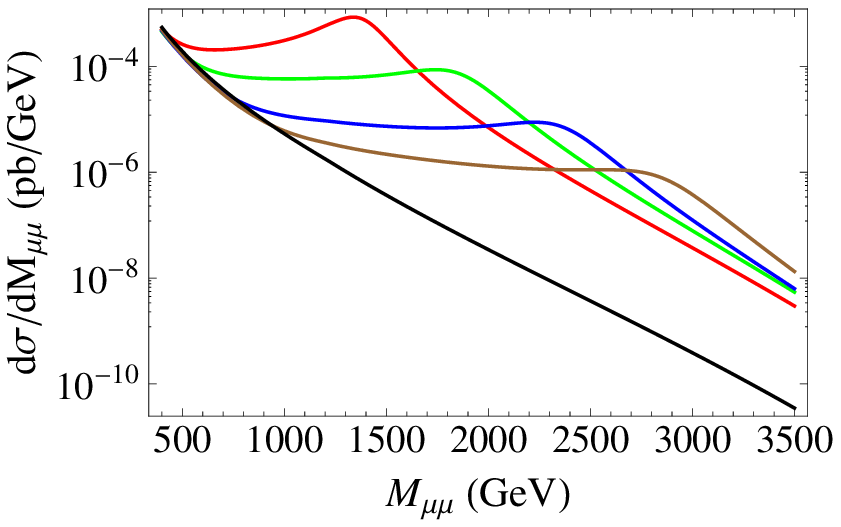}
\caption{Differential cross sections for RS (top) and non-universal $Z'$ (bottom),
as defined in section \ref{sec:models}. Top left: 
graviton of mass 2 TeV with $k/{\overline M_{\mathrm{Pl}}}$ = 0.7
(highest), 0.5, 0.3, and 0.1 (lowest).  
Top right: $k/{\overline M_{\mathrm{Pl}}}=0.4$
and masses 1.4 TeV (left), 1.9 TeV, 2.4 TeV, and 2.9 TeV (right). 
Bottom left: $Z'$ of mass 2 TeV with $\epsilon$ = 1.4 (highest), 1.1, 0.8, and 0.5 (lowest).  
Bottom right: $Z'$ with $\epsilon=1$ and masses 1.4 TeV (left), 1.9 TeV, 2.4 TeV, and 2.9 TeV (right).
The SM is shown in black on all plots.}
\label{fig:InvDist}
\end{center}
\end{figure}

\begin{figure}[!b]
\begin{center}
\includegraphics[scale=0.9]{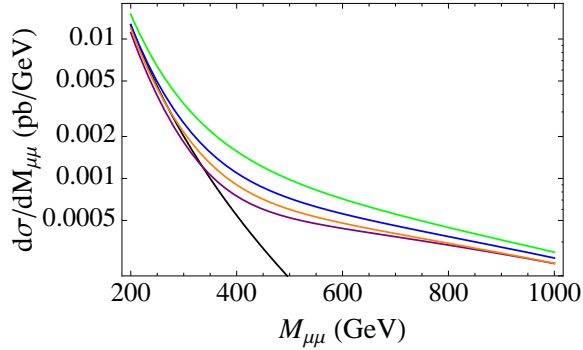}
\caption{Differential cross sections for four different contact interactions at 
$\sqrt s = 7\,\,\mathrm{TeV}$.  The models are given by Eq.~(\ref{eq:contact}) with various 
signs of $\eta$.  From highest to lowest:
VV constructive (green), LL constructive (blue),
LL destructive (orange), VV destructive (purple), SM (black).  The contact scale is 
$\Lambda = 4\,\,\mathrm{TeV}$ for the VV models and $\Lambda = 2.87\,\,\mathrm{TeV}$ for the LL models.
The energy scales are chosen to have the same high-energy behavior for all four.}
\label{fig:Contact}
\end{center}
\end{figure}

As the coupling of the theory increases, the resulting resonance broadens and flattens,
losing its distinctive ``bump'', and instead begins to look more and more like background,
a contact interaction, or some other new physics that predicts excess events at high
invariant mass.  Nonetheless, discriminating a resonance from other possible models
(including the SM)  should be possible
for many of the broad resonance models we study, although it is not
\emph{a priori} obvious that this is the case.
To show why  model discrimination may be difficult,
we plot the differential cross section for both RS
and $Z'$ resonances in Fig.~\ref{fig:InvDist},
varying the mass and coupling to get a qualitative sense of how the
invariant mass distributions
change with these parameters.  We also plot in Fig.~\ref{fig:Contact}
 the differential cross section for a set of 
contact interactions for comparison with the resonances.

\begin{figure}[t]
\begin{center}
\includegraphics[scale=0.27,angle=270]{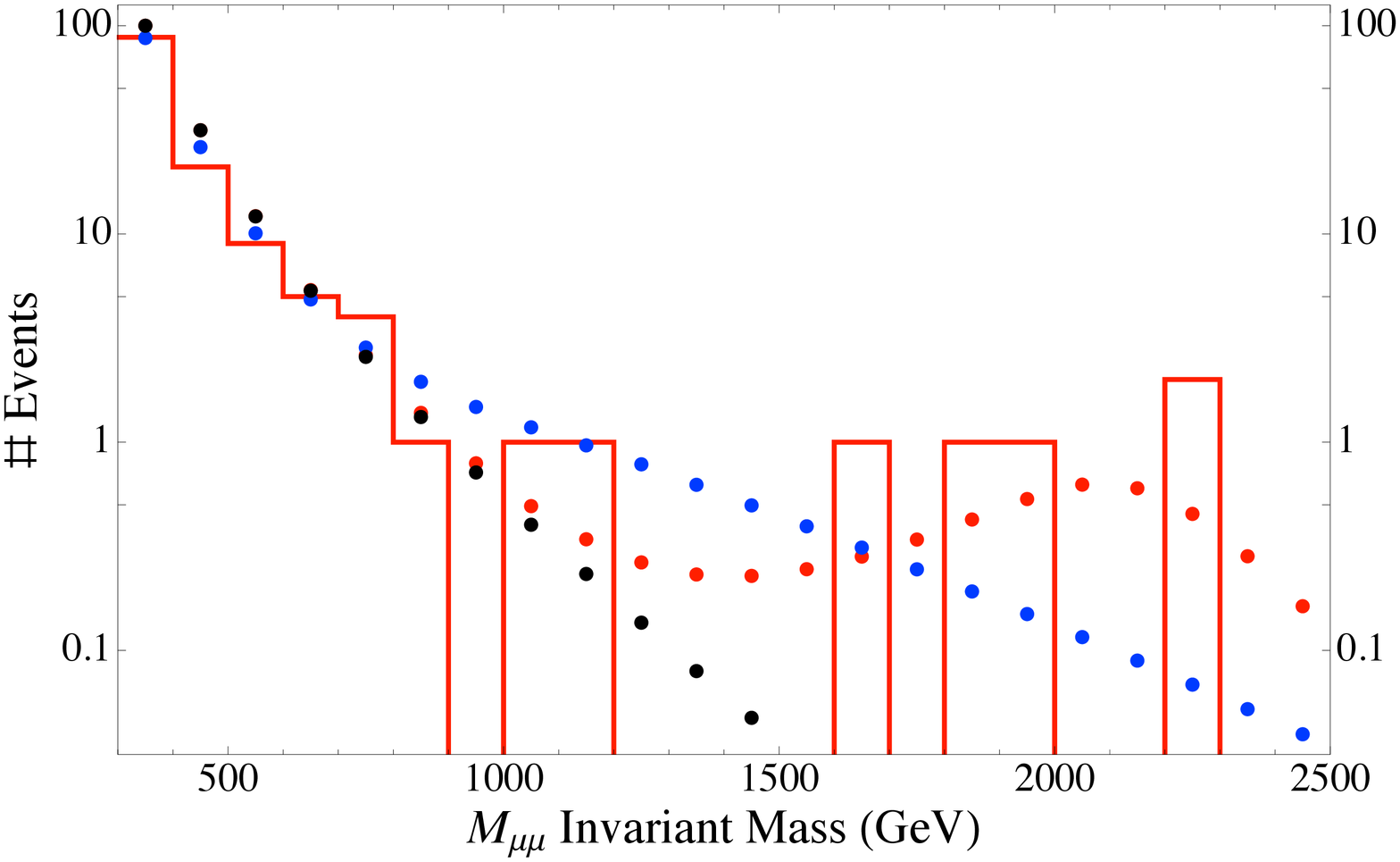} 
\includegraphics[scale=0.27,angle=270]{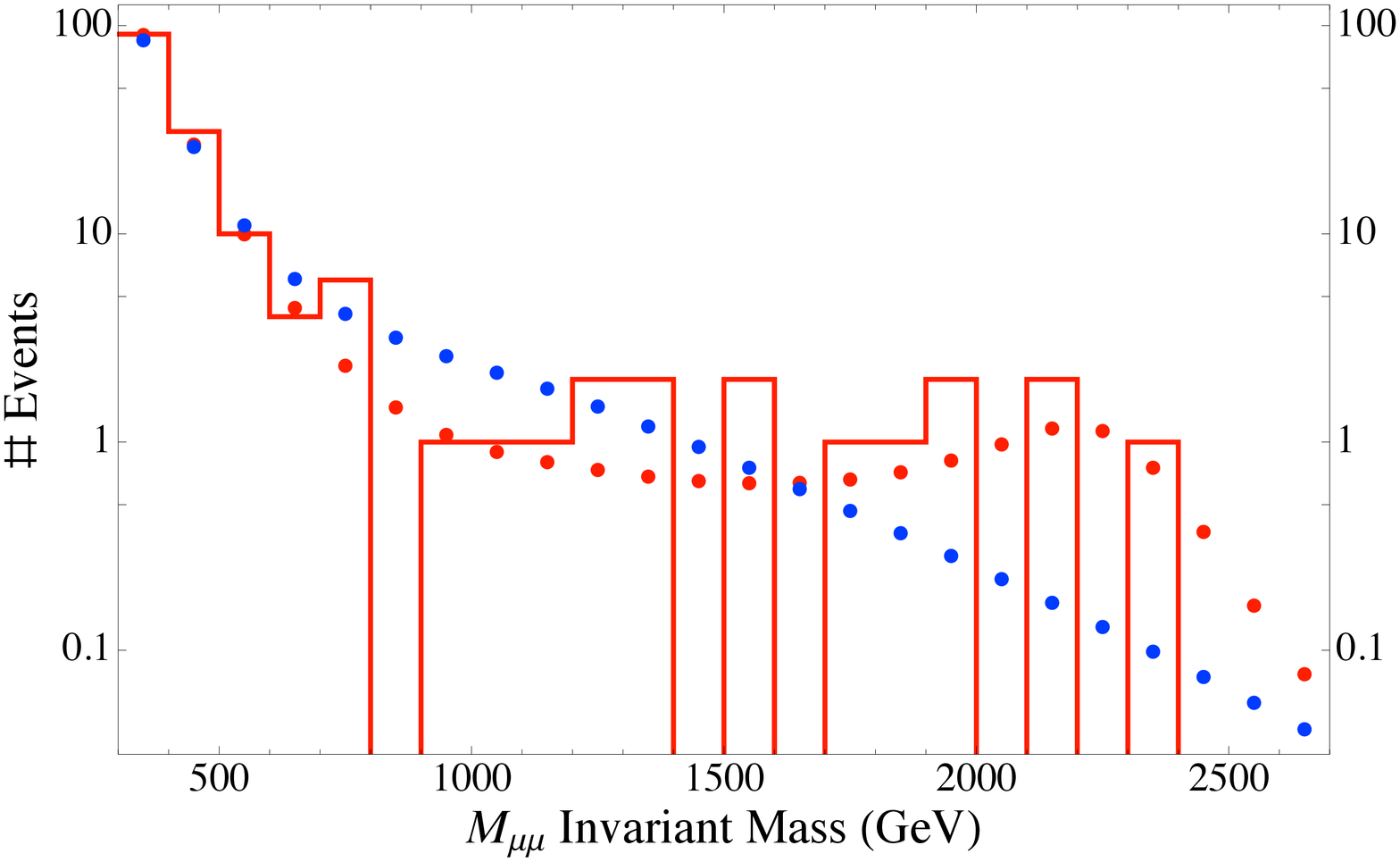}
\caption{Illustration of the potential difficulty of determining new model type from data, 
whether it be a resonance or a contact interaction.  Data was simulated for RS (left) and 
non-universal $Z'$ (right), and is shown in blocked red.  Best fits to data from resonance and 
contact models are shown in red and blue dots, respectively.  Black dots show the SM prediction.}
\label{fig:ZPfake}
\end{center}
\end{figure}

To illustrate the aforementioned difficulties
in distinguishing different models of new physics through the
invariant mass distribution, we ran two simulations showing possible outcomes of an experiment
at $\sqrt s = 7\,\,\mathrm{TeV}$ and $1\,\,\mathrm{fb}^{-1}$ (see Fig.~\ref{fig:ZPfake}). 
We fit a resonance and a contact interaction to the data and, by sight, it seems that both
might be consistent with the data up to statistical fluctuations.  We want to see when it is 
possible to reliably distinguish between the two types of new physics, and how this discriminating
capability varies with the parameters of the theory.\footnote{If we take the 
resonance mass and coupling to be very large, we can integrate out the new physics
and we \emph{do} end up
with a contact interaction.  This, however, is not what we're talking about here; rather, we are
concerned about the case where the observed signal events are in the region around the resonance
peak.}  In section \ref{sec:discrimination},
we will look at two methods for doing so.
The first is relatively simple, looking for an absolute rise
in the differential cross section.  The second employs a
statistic that is commonly used in experiments with few data points
in each bin. We show that
both of these methods can be used to distinguish
resonances from contact interactions for most models predicting
$>5$ events in the early LHC running, though the more careful statistical method covers a
larger range of parameters.


\section{Models of new physics} \label{sec:models}

In this section we define the two types of resonance models we use to demonstrate our methods, namely $Z'$ models of a new $U(1)'$ gauge boson,
as well as RS models with at least one accessible KK graviton resonance.
We also define the contact interaction models that we will compare with the resonances.

\subsection{$Z^{\prime}$ models}

$Z'$ bosons   commonly
appear in physics beyond the SM
(for a recent review, see \cite{Langacker:2008yv}).
The coupling of a $Z'$ to the SM can be written as
\begin{equation}
\mathcal{L}_{Z'} = g_{Z'}\,Z_{\mu}'J_{Z'}^{\mu},
\end{equation}
where
\begin{equation}
J_{Z'}^{\mu} = \sum_f Q(f)\,\bar{f}\gamma^{\mu}f.\textit{}
\end{equation}
The charges $Q(f)$ will be discussed more in-depth shortly,
although following \cite{Salvioni:2009jp},
we will focus on $Z'$ that couple to $B-L$ and
$B-3L_{\mu}$.  
We define the $g_{Z'}$ coupling in terms of
the $g_Z$ coupling by introducing the parameter $\epsilon$:
\begin{equation}
\epsilon = \frac{g_{Z'}}{g_Z}.
\end{equation}

\begin{figure}[t]
\begin{center}
\includegraphics[scale=1]{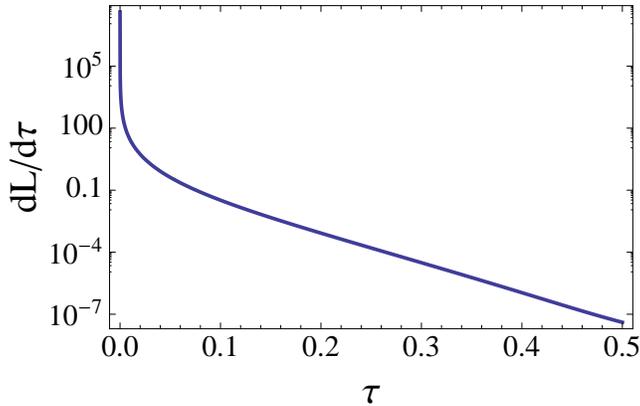}
\caption{Parton luminosity $d\mathcal L_u/d\tau$ ($\tau=\hat s/s$)
for $u\bar u$ contribution to scattering processes at the LHC,
$\sqrt s = 7\,\,\mathrm{TeV}$.}
\label{fig:Luminosity}
\end{center}
\end{figure}

We compute the differential and total cross sections to determine how these
quantities scale with the model parameters.  The partonic cross section has the form
\begin{equation}
\hat{\sigma} \sim \epsilon^4\frac{\hat s}{(\hat s-M_{Z'}^2)^2+\mathrm{Im}\,\Pi(\hat s)^2}+\mathrm{interference},
\end{equation}
where near the resonance or in the narrow width approximation, we have a partial width to states of
mass $m$
\begin{equation}
\Gamma\sim M_{Z'}\,\epsilon^2\left[1+\mathcal{O}\left(\frac{m^2}{M_{Z'}^2}\right)\right].
\end{equation}
The partonic cross section evaluated at the pole is
\begin{equation}
\hat{\sigma}(M_{\mathrm g}^2)\sim\frac{1}{M_{Z'}^2},
\end{equation}
and so the peak cross section decreases with higher $M_{Z'}$, as expected.  In a hadron collider,
however,  the partonic cross section, $\hat{ \sigma}(\hat{s};  ij \to \ell^+ \ell^-) 
\equiv \hat{\sigma}_{ij}(\hat{s})$, is integrated against the 
corresponding parton luminosity function, $d \mathcal{L}_{ij} / d \tau$, defined as follows: 
\begin{align}
\frac{ d\mathcal{L}_{ij} }{d \tau} (\tau ) 
=
    \frac{1}{1 + \delta_{ij}} \int_\tau^1  \frac{dx}{x }  
    [ f_{i}(x)f_{j}(\tau/x) + f_{i}(\tau/x)f_{j}(x) ],   
\end{align}
where $\tau = \hat{s}/s$, to give the hadronic cross section 
\begin{align}
\sigma^{\textrm{hadronic}} ( s )  = 
               \sum_{ij}
               \int_0^1 d\tau 
               \frac{ d\mathcal{L}_{ij} }{d \tau}
               \hat{\sigma}_{ij}( \tau s ).
\end{align}
The hadronic cross section can  be evaluated analytically only in the narrow width approximation,
giving for CM energy  $\sqrt s$
\begin{equation}
\sigma \sim \frac{\epsilon^2}{s}  \cdot\frac{d\mathcal{L} }{d\tau}(\tau_{Z'}),
\end{equation}
where $\tau_{Z'} \equiv M_{Z'}^2/ s$.
This is independent of the mass $M_{Z'}$, except for the dependence on $\tau_{\mathrm Z'}$
in the parton luminosity function.  The parton luminosity function decreases monotonically
with $\tau$, as seen in Fig.~\ref{fig:Luminosity}.\footnote{We probe the regime where the luminosity
can no longer be taken as a simple power law: the total 
partonic luminosity from $u\bar u$ collisions at the LHC
is approximately given by the expression
$d\mathcal{L}/d\tau=0.036 \tau^{-2.04}(1-\tau)^{0.02}\exp(-27.2\tau)$.}

$Z'$ phenomenology is dictated by the parameters $M_{Z'}$ and $\epsilon$.
In our analysis, we consider all values of these parameters consistent with
experimental constraints.
To observe higher-mass resonances,   the coupling $\epsilon$
must be sufficiently large to compensate for the
small parton luminosity, meaning that the $Z'$ models most visible during the early
LHC running will be broad.\footnote{Strictly speaking, this scaling argument only applies
in the narrow
width approximation, but we find a similar approximate
power law scaling of hadronic cross section with
$\epsilon$ even at large coupling, and our conclusion remains the same.}

  The only rigorous constraint on $Z'$ couplings comes from anomaly cancelation requirements.
These can be satisfied in myriad ways - for instance, by the presence of
hidden exotic states - but the simplest are ``minimal models'', which consist purely of couplings
to the SM plus right-handed neutrinos \cite{Salvioni:2009mt,Basso:2010pe}. 
The minimal models allow us to adequately study the various $Z'$ possibilities in a systematic way
without introducing too many additional parameters,
and we will restrict our attention to these.  Anomaly cancelation then limits the overall charges 
to a linear combination of two independent charge assignments: hypercharge and $B-L$.  
The hypercharge component leads to mixing with the $Z$ and is tightly constrained by LEP measurements.
In order to study $Z'$ models with stronger coupling to the SM,
we will focus on $Z'$ of the $B-L$ type. Strictly speaking, the couplings will mix under
renormalization group running \cite{Salvioni:2009mt}, so $Z'$ models
that unify with the SM at the GUT scale will generically
see both types of charges, but this effect is small and, for the purposes of LHC phenomenology,
we can ignore  higher energies and simply examine the couplings at the TeV scale.

Although not as tightly constrained as a $Z'$ that couples to hypercharge, $B-L$ models are also strongly
constrained by electroweak data if they are flavor-universal due to the constraints on
electron-quark 4-fermion operators from LEP.  If we only
consider flavor-universal models, the LEP constraints are sufficiently strong to exclude all broad
resonances with $\mathcal O(\mathrm{TeV)}$ mass, and since these are the only type that will be visible
in early LHC running,
there will be no new parameter space open for exploration.  A way around this is to couple the $Z'$
exclusively to muons in the leptonic sector \cite{Salvioni:2009jp,Davidson:2001ji};
the theory will still be anomaly-free if the charges are $B-3L_{\mu}$.  These non-universal 
models evade the bounds from LEP while giving large branching fraction to muon final states, 
enhancing the visible signal at the LHC.  This means that there are regions of the parameter
space that are accessible at the early LHC but that
have not already been excluded at LEP.  Our analysis generalizes that of \cite{Salvioni:2009jp}
 pertaining to $Z'$ coupling to $B-3L_{\mu}$,
considering in particular the case of broad resonances and its experimental implications.


\subsection{Brane RS models}

Many of the best-motivated theories of new physics are those providing a solution to the hierarchy
problem.  One class of examples are RS models, in which the hierarchy
between the Planck and electroweak scales is established with the addition of
a warped extra dimension and the Higgs field confined to a brane at the IR scale \cite{Randall:1999ee}.
For theories with the entire SM confined to the TeV brane,   KK modes
of the graviton could be detectable in early running if the TeV-brane energy scale is sufficiently low.

The couplings of higher KK modes of the graviton are suppressed by the warped scale
$\Lambda_{\pi} = \overline M_{\mathrm{Pl}}\,e^{-\pi\,k\,r_{\mathrm c}}$.  
Here, $k$ is the AdS curvature
scale and $r_{\mathrm c}$ is the radius of the compact dimension.
If the hierarchy problem is solved, then $\Lambda_{\pi}\sim\mathrm{TeV}$.  
The Lagrangian for the coupling of the KK graviton modes is
\begin{equation}
\mathcal{L} = -\frac{1}{\Lambda_{\pi}}\sum_{n=1}^{\infty}T^{\mu\nu}h_{\mu\nu}^{(n)},
\end{equation}
with energy-momentum tensor $T^{\mu\nu}$ and $n$-th KK graviton $h^{(n)}_{\mu\nu}$.  The free parameters in this model
can be taken to be $k$ and $\Lambda_{\pi}$; for phenomenological reasons, it is more convenient
to exchange these for the dimensionless parameter $k/\overline M_{\mathrm{Pl}}$ and the first KK graviton mass
\begin{equation}
M_{\mathrm g} = x_1\,\Lambda_{\pi}\left(\frac{k}{\overline M_{\mathrm{Pl}}}\right),
\end{equation}
where $x_1 = 3.83171$ is the first zero of the Bessel function $J_1(x)$.
The subscript $g$ is used for the first
mode because it is the graviton mode that will first appear at the LHC,
and higher modes will likely be too heavy for discovery in the early running.
These parameters are better-motivated
from the  standpoint of a resonance search
and are directly analogous to the parameters $\epsilon$ and $M_{Z'}$ in $Z'$ models.

As with the $Z'$, we compute the partonic and hadronic cross sections
for lepton production--here via KK graviton exchange.  For both gluon- and quark-initiated
processes, the partonic cross section takes the form
\begin{equation}
\hat\sigma \sim \frac{\hat s^3}{\Lambda_{\pi}^4}\cdot\frac{1}{(\hat s-M_{\mathrm g}^2)^2+\mathrm{Im}\,\Pi(\hat s)^2},
\end{equation}
where $\Pi(\hat s)$ is the graviton self-energy.  On-resonance, or in the narrow width approximation, 
the self-energy contribution can be simplified to give
\begin{equation}
\mathrm{Im}\,\Pi(\hat s) \approx \mathrm{Im}\,\Pi(M_{\mathrm g}^2) = M_{\mathrm g}\,\Gamma,
\end{equation}
where the graviton partial width to final states of mass $m$ scales as
\begin{equation}
\Gamma\sim M_{\mathrm g}\left(\frac{k}{M_{\mathrm{Pl}}}\right)^2\left[1+\mathcal{O}\left(\frac{m^2}{M_{\mathrm g}^2}\right)\right].
\end{equation}
 
The partonic cross section evaluated at the pole is
\begin{equation}
\hat{\sigma}(M_{\mathrm g}^2)\sim\frac{1}{M_{\mathrm g}^2},
\end{equation}
and the hadronic cross section is
\begin{equation}
\sigma \sim \frac{\left(k/\overline M_{\mathrm{Pl}}\right)^2}{s}  \cdot\frac{d\mathcal{L} }{d\tau}(\tau_{\mathrm g}),
\end{equation}
where $\tau_{\mathrm g} \equiv M_{\mathrm g}^2/ s$.
This behavior is clearly analogous to that for a $Z'$, and we are likewise led to the conclusion
that strongly-coupled physics and broad resonances will appear first at the LHC.

Many analyses, beginning with \cite{Davoudiasl:2000wi},  consider only RS models with coupling $k/\overline M_{\mathrm{Pl}}<0.1$
due to perturbativity constraints,
restricting their studies to narrow resonances.  This constraint, which comes from bounding the tree-level AdS curvature $|R_5| = 20\,k^2<M^2$, neglects loop effects and is likely overly conservative.  

\subsection{Contact interactions} \label{sec:contact}
We now discuss contact interactions in more detail, as we saw in 
section \ref{sec:resorcontact} that, with small statistics,
they might be confused with broad resonances.
Contact interactions are convenient for parameterizing
the low-energy behavior of new physics contributions.
We consider generic contact interactions that are consistent 
with current experimental data and remain agnostic about
possible UV completions of the theory.

The contact interactions most relevant to us are those involving
quarks and muons. They are defined by a scale $\Lambda$, 
and we use the parameterization \cite{Eichten:1984eu}
\begin{align} 
\label{eq:contact}
\mathcal{L} 
    = \frac{4\pi}{\Lambda^2}
    \Bigl[
    &
        \eta_{\mathrm{LL}}
        \left(\bar{q}_{  \mathrm L}\gamma^{\nu}q_{  \mathrm L}\right)
        \left(\bar{\mu}_{\mathrm L}\gamma_{\nu}\mu_{\mathrm L}\right)
        +
        \eta_{\mathrm{LR}}
        \left(\bar{q}_{  \mathrm L}\gamma^{\nu}q_{  \mathrm L}\right)
        \left(\bar{\mu}_{\mathrm R}\gamma_{\nu}\mu_{\mathrm R}\right)
    \nonumber \\
    &
        +
        \eta_{\mathrm{RL}}
        \left(\bar{q}_{  \mathrm R}\gamma^{\nu}q_{  \mathrm R}\right)
        \left(\bar{\mu}_{\mathrm L}\gamma_{\nu}\mu_{\mathrm L}\right)
        +
        \eta_{\mathrm{RR}}
        \left(\bar{q}_{  \mathrm R}\gamma^{\nu}q_{  \mathrm R}\right)
        \left(\bar{\mu}_{\mathrm R}\gamma_{\nu}\mu_{\mathrm R}\right)
    \Bigr].
\end{align}
where $|\eta|=1$ or 0 and can give chiral or axial/vector-like couplings.

Models with a vector-like coupling, denoted by VV, will have all couplings
set to 1 or $-1$,
each with the same sign. 
A left-handed model, denoted by LL, has only one non-zero coupling, $\eta_{\mathrm{LL}}$, and thus only
couples to left (right) handed quarks (anti-quarks) and muons (anti-muons). 
For these two models, the signs of the $\eta$'s determine the nature of the 
interference with the SM.  The SM is recovered in the limit $\Lambda\rightarrow\infty$. 

The contact interaction (\ref{eq:contact}) arises from integrating out some
spin-1 degrees of freedom.  This seems like the best choice for mimicking
a $Z'$ resonance, but at first glance, it may seem that a contact interaction
generated by integrating out spin-2 degrees of freedom might do better at faking
a graviton resonance. Such contact interactions, however, are dimension 8
and suppressed by extra powers of the scale $\Lambda$.  As a result, the event
rate from such operators will not compete with the event rate from a broad
resonance without running into serious conflict with existing
experimental bounds.  Thus, we find that even for RS gravitons, the 
interaction (\ref{eq:contact}) is the best basis for comparison (we may also
wish to consider scalar-type couplings, but as in the high energy limit,
this only affects the angular distribution, we do not consider them here).


\section{LHC dimuon cross sections} \label{sec:xsec}
The current goal for early LHC running is $1\,\,\mathrm{fb}^{-1}$ at $\sqrt s = 7\,\,\mathrm{TeV}$.  
To get  a sense of the ultimate LHC reach, however, and also to see qualitatively what the possible
outcomes of resonance searches may be at later times,
we consider both the early run and also the cross sections at $\sqrt
s = 10\,\,\mathrm{TeV}$ and $\sqrt s = 14\,\,\mathrm{TeV}$, even though the analyses presented in
this paper are mostly concerned with the first case.  We consider the RS and 
$Z'$ models discussed in section \ref{sec:models} for different values of the mass
and coupling.  As anticipated in section \ref{sec:resorcontact}, we demonstrate that the 
largest cross sections are in regions of parameter space with broad resonances, leading
to the question of model discrimination between the broad resonance and a contact
interaction that will be addressed in section \ref{sec:discrimination}.

\begin{figure}[h]
\begin{center}
\includegraphics[scale=0.75]{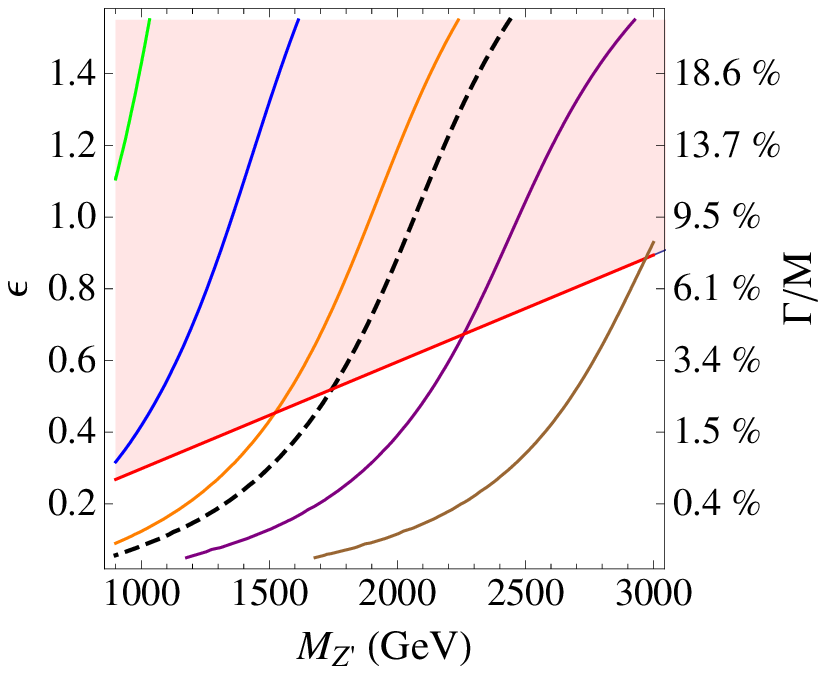}
\includegraphics[scale=0.75]{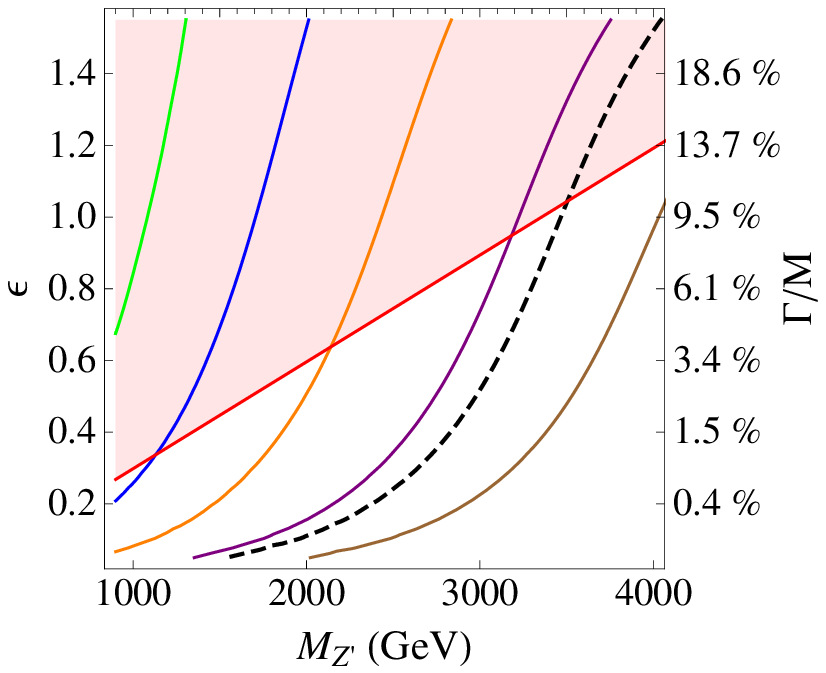}
\includegraphics[scale=0.75]{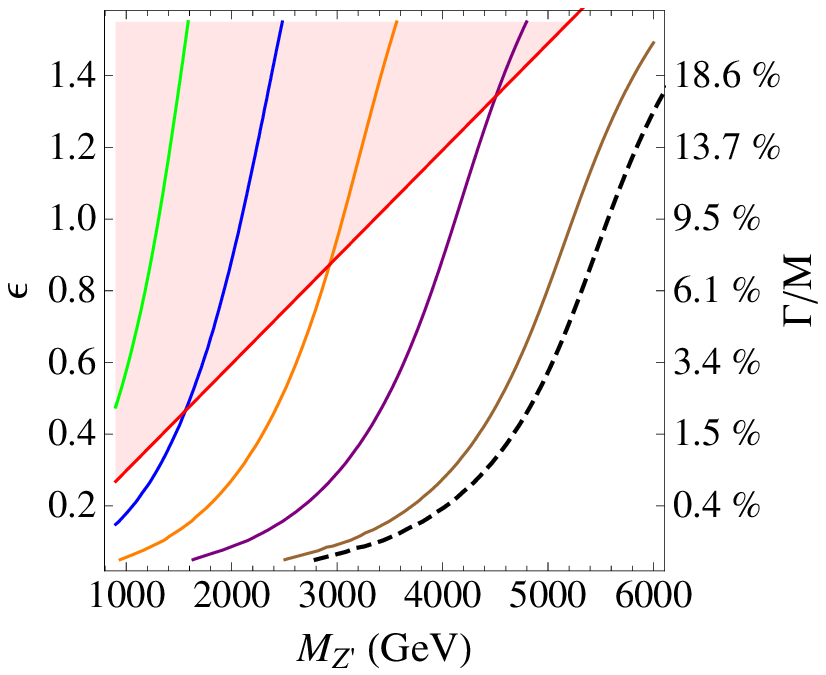}
\caption{Plots showing cross sections to dimuon final states
for a $Z'$ with $B-L$ charges to SM. 
The cross sections are shown for $\sqrt s=7\,\,\mathrm{TeV}$ (above left), 10 TeV
(above center), and 14 TeV (below).
The solid line cross section contours are, from top to bottom: 1 pb (green),
100 fb, 10 fb, 1 fb, and 100 ab (brown).
Dashed curves indicate the cross section for 5 events at certain benchmark luminosities:
1 $\mathrm{fb}^{-1}$ at 7 TeV, 10 $\mathrm{fb}^{-1}$ at 10 TeV, and 100 $\mathrm{fb}^{-1}$ at 14 TeV. 
The shaded region is excluded by LEP and the Tevatron.  }
\label{fig:BMLxsec}
\end{center}
\end{figure}

\begin{figure}[h]
\begin{center}
\includegraphics[scale=0.75]{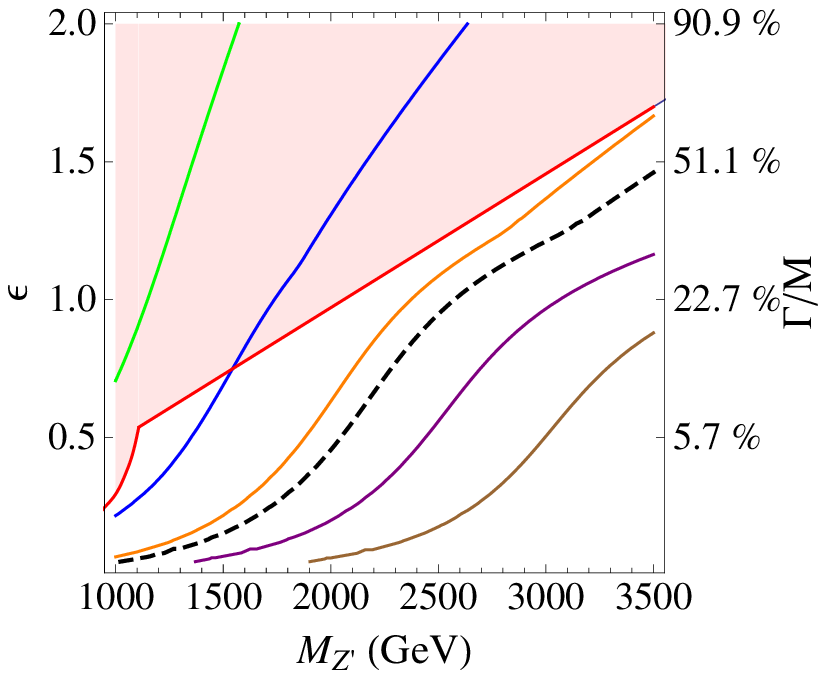}
\includegraphics[scale=0.75]{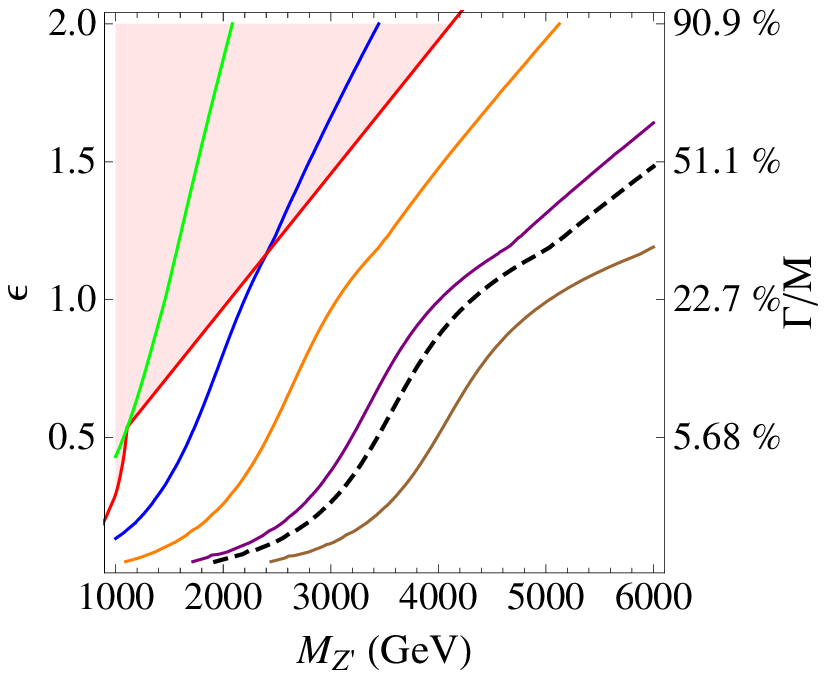}
\includegraphics[scale=0.75]{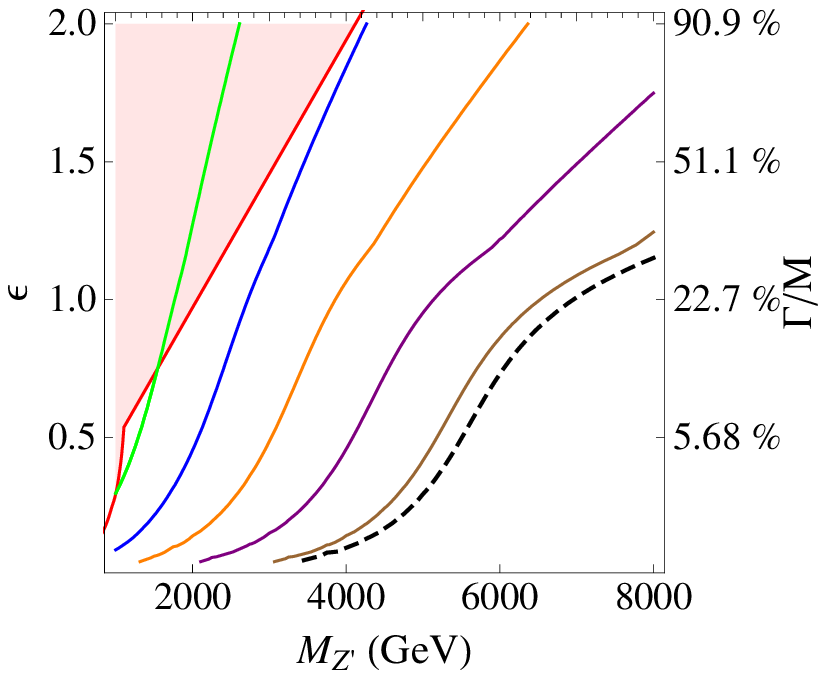}
\caption{Plots showing cross sections to dimuon final states
for a $Z'$ with $B-3L_{\mu}$ charges to SM. 
The cross sections are shown for $\sqrt s=7\,\,\mathrm{TeV}$ (above left), 10 TeV
(above center), and 14 TeV (below).
The solid line cross section contours are, from top to bottom: 1 pb (green),
100 fb, 10 fb, 1 fb, and 100 ab (brown).
Dashed curves indicate the cross section for 5 events at certain benchmark luminosities:
1 $\mathrm{fb}^{-1}$ at 7 TeV, 10 $\mathrm{fb}^{-1}$ at 10 TeV, and 100 $\mathrm{fb}^{-1}$ at 14 TeV.  
The shaded region is excluded by LEP and the Tevatron. }
\label{fig:NUxsec}
\end{center}
\end{figure}

\begin{figure}[t]
\begin{center}
\includegraphics[scale=0.7]{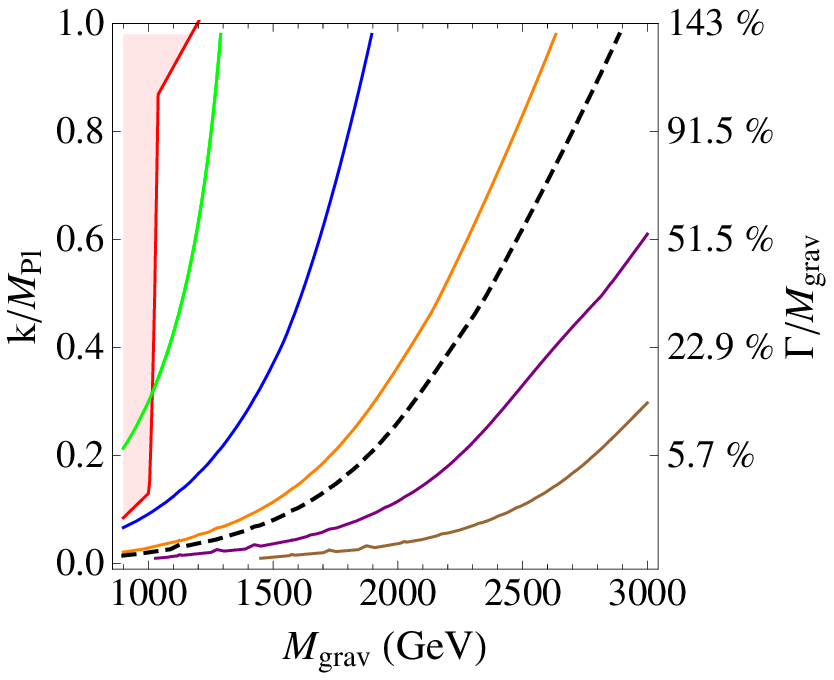}
\includegraphics[scale=0.7]{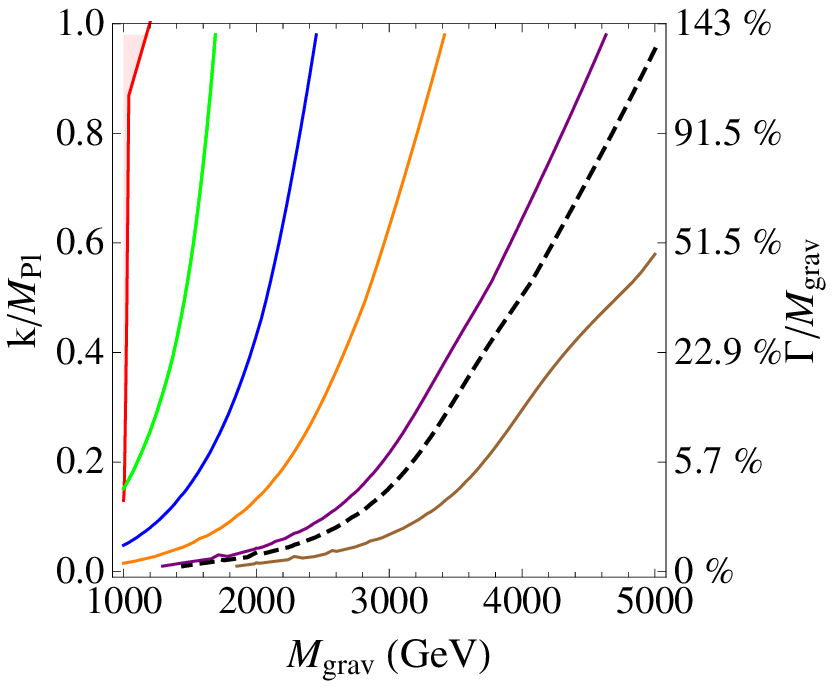}
\includegraphics[scale=0.7]{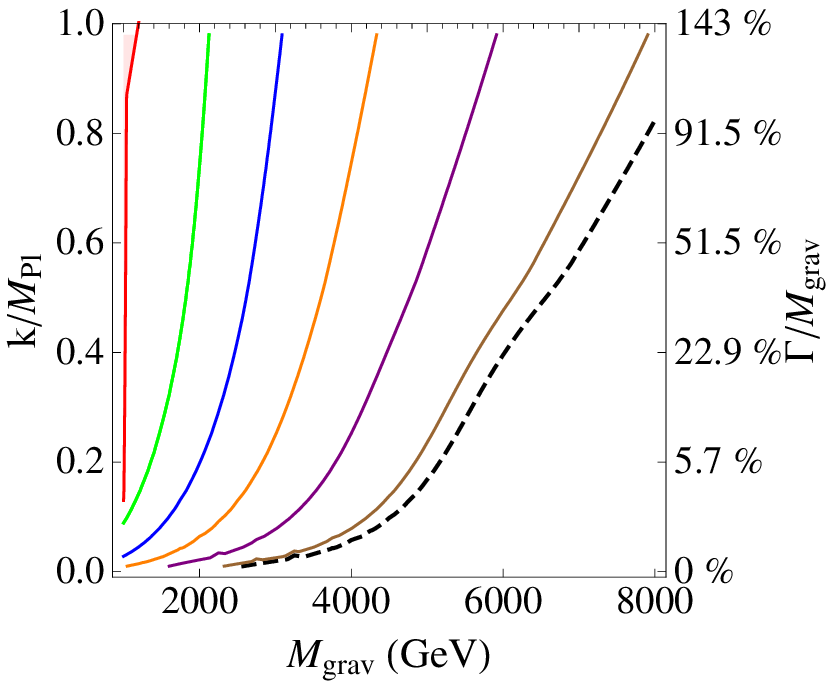}
\caption{Plots showing cross sections to dimuon final states
forfor the first KK mode of the graviton in
RS models.
The cross sections are shown for $\sqrt s=7\,\,\mathrm{TeV}$ (above left), 10 TeV
(above center), and 14 TeV (below).
The solid line cross section contours are, from top to bottom: 1 pb (green),
100 fb, 10 fb, 1 fb, and 100 ab (brown).
Dashed curves indicate the cross section for 5 events at certain benchmark luminosities:
1 $\mathrm{fb}^{-1}$ at 7 TeV, 10 $\mathrm{fb}^{-1}$ at 10 TeV, and 100 $\mathrm{fb}^{-1}$ at 14 TeV. 
The shaded region is excluded by LEP and the Tevatron.  }
\label{fig:RSxsec}
\end{center}
\end{figure}

We present our results for
$Z'$ signal cross sections
 in Fig.~\ref{fig:BMLxsec} and \ref{fig:NUxsec}
 (for the flavor universal and non-universal
 cases, respectively), and our results for RS signal cross sections
in Fig.~\ref{fig:RSxsec}. We mark contours of constant
cross section, including the contours that will give 5 events at certain benchmark luminosities:
$1\,\,\mathrm{fb}^{-1}$ at 7 TeV, $10\,\,\mathrm{fb}^{-1}$ at 10 TeV, and $100\,\,\mathrm{fb}^{-1}$
at 14 TeV.  We indicate constraints using dashed lines.  Details of the calculation and the
relevant direct and indirect constraints follow later in this section. 
In the plots shown here, the regions of parameter space most ripe for exploration in
the Drell-Yan channel at the LHC are those that couple strongly to muons, giving rise to broad
resonances.  The only exceptions are the $Z'$ models that are pure $B-L$: since they couple universally
to muons and electrons, the broad resonance regions are largely excluded by LEP.  Indeed, even at
high CM energy and luminosity, a $B-L$ $Z'$ will appear as a narrow resonance and the
standard search strategies will apply.


We compute the differential cross sections at 
leading order in $\alpha$
using the MSTW 2008 PDFs \cite{Martin:2009iq}, with invariant-mass-dependent K-factors
applied to quantify NLO QCD effects \cite{ATLAS:expected,Li:2006yv}.  Other necessary cuts
due to detector geometry or triggering,
for example $|\eta|<2.5$ and $p_{\mathrm T}>20\,\,\mathrm{GeV}$, are incorporated into the
computation by employing an invariant-mass-dependent acceptance, following
\cite{Salvioni:2009mt,CMS:muon}.  
The total cross section is then found by integrating the invariant mass distribution
in some window.
Typically, the width of this window is some multiple of the resonance width $\Gamma$, but
when we look at very broad resonances with $M\sim\Gamma$, this will include events at very low
invariant mass where SM background dominates.  We therefore instead choose a lower cut-off for our window
of $\mathrm{Max}(M-2\Gamma,M_*)$, where $M_*$ is the invariant mass at which
the signal differential
cross section is twice as large than the background.  Our upper cut-off for the window of
integration will be $\sqrt s$, the CM energy of the collider.

  We do not exclude any regions
of parameter space solely for reasons of theoretical prejudice, but consider only experimental constraints.
The most important ones  for our study
will be the bounds on 4-fermion operators from LEP \cite{Alcaraz:2006mx}
and the Tevatron \cite{Titov:2005hd}.
The bounds from LEP in particular strongly constrain new physics coupling to electrons, such 
as the $Z'$ coupling to $B-L$.  One exception is the RS KK graviton that we have been discussing.
It couples to the energy-momentum tensor and the lowest-dimension operator contributing to
$s$-channel graviton exchange is dimension 8, which is much less strongly constrained than
the dimension-6 operators generated by massive vector exchange.

We also consider the strongest direct bounds on neutral
resonances, which come from the Tevatron \cite{Abazov:2010xh,Aaltonen:2008ah}.  
These direct bounds do not extend much beyond 1 TeV because the Tevatron parton luminosity 
falls off very rapidly above this scale.  Finally, we incoporate the bounds from muon $g-2$ experiments
\cite{Kim:2001rc,Park:2001uc,Amsler:2008zzb}, although these do not strongly constrain 
RS models due to the cut-off dependence of the graviton contribution to $g-2$, and so there are many
values of $M_{\mathrm g}$ and $k/\overline M_{\mathrm{Pl}}$ that can give the correct contribution.  For $Z'$
models, the constraints from 4-fermion operators are stronger than those from $g-2$.


\section{Shape discrimination}\label{sec:discrimination}

\subsection{Upturn analysis}
As a simple first attempt at studying the shape of the distribution, we take advantage of one of the
defining characteristics of a resonance: namely, the accompanying ``bump'', or absolute rise and
fall in the cross section.  The background (or a contact interaction)
will never actually give an increase in the cross section. \textit{}
We therefore look at regions of parameter space where we can, with $5\sigma$ confidence, say that
there is a rise in the cross section over a flat background, indicative of a resonance with 
reasonable certainty.   
We find that for both KK gravitons and $Z'$, the upturn analysis works for masses up to
1.5-2 TeV, and widths up to about 40\% for $Z'$ and 90\% for RS, depending on the mass.

\begin{figure}[t]
\begin{center}
\includegraphics[scale=0.9]{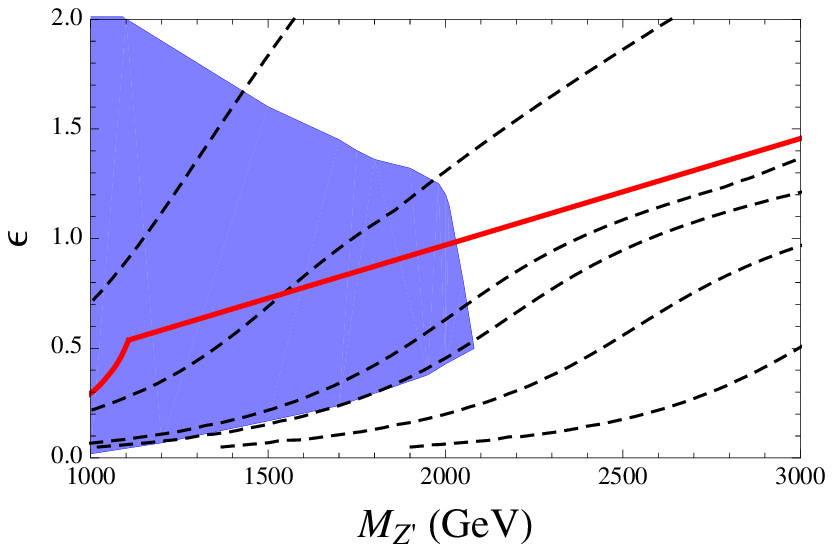}
\includegraphics[scale=0.9]{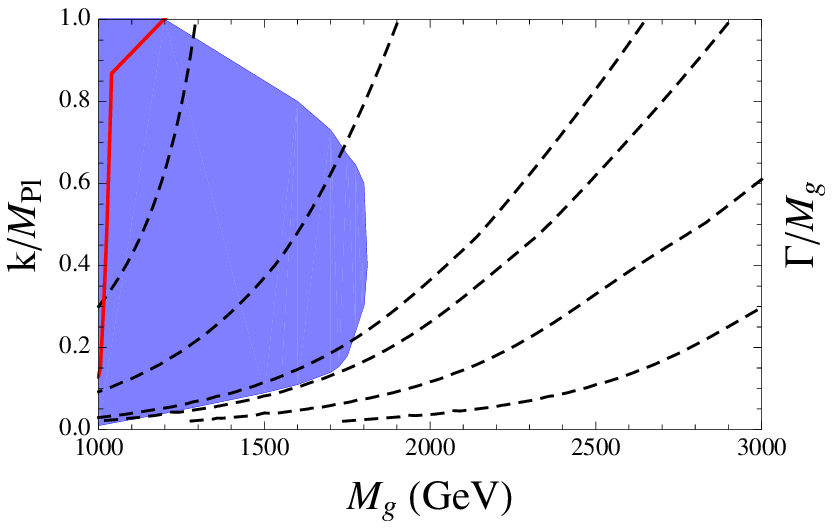}
\caption{Regions of parameter space for non-universal $Z'$ (left) and RS (right) models 
accessible ($5\sigma$) at the early LHC via the upturn analysis are shaded in blue. 
Cross-section contours are shown in dashed black lines: 1 pb (top),
100 fb, 10 fb, 5 fb, 1 fb,
and 100 ab (bottom).
Excluded regions are to the left of the solid red line.}
\label{fig:RSupturn}
\end{center}
\end{figure}

To determine the reach of this analysis,
we search for the local minimum of the differential cross section, as well as the
corresponding point on the other side of the putative resonance peak.
We then compute the integrated cross section in this region 
and compare it to a conjectured flat background.  
If we see a $5\sigma$ excess of events, then we know to a very high degree of confidence that 
we are seeing something new, since nowhere in the SM do we expect to see such behavior.  
We present our results for the upturn analysis
at $1\,\,\mathrm{fb}^{-1}$ at the 7 TeV LHC
in Fig.~\ref{fig:RSupturn}.

The behavior is as expected for narrow resonances, where nearly all of the signal events are above
the upturn in cross section, and so the upturn analysis works everywhere that a normal
counting experiment would.  For very broad resonances (this definition 
depends on the precise model and resonance mass,
but typically  involves a ratio of width to mass that is larger than 50\% for RS and
30\% for $Z'$),
the convolution with the falling parton luminosity function results in fewer events above the upturn
(and, in some cases, no upturn at all), and they cannot be reached by this analysis.

This method is quick and easy but is overly conservative, since
the background (and other, similar-looking types of new
physics, such as contact interactions) are falling and not constant.  Therefore,
we are neglecting  some
broad resonances that should be distinguishable from background
because they don't show a particularly pronounced upturn in
the differential cross section.  We expect that the LHC reach for broad
resonances should be  more extensive than that shown here.
Another shortcoming of this analysis is that it fails to account for
statistical fluctuations in determining the start and endpoints for the integration window.  
An upturn could just as well come from a downward fluctuation
in the lower-invariant-mass region, and our
analysis does not consider this possibility.


\subsection{Maximum likelihood method}\label{sec:mlm}
We now present a more powerful method for discriminating resonances from contact
interactions than the analysis presented above.  We perform a full maximum likelihood
anlysis \cite{Amsler:2008zzb} on the invariant mass distribution
from a $1\,\,\mathrm{fb}^{-1}$
experiment.  We simulate experiments assuming an
underlying resonance as new physics,
compare the maximum likelihood ratio 
obtained from resonance
and contact interaction fits to the data, and determine over what
parameters we can distinguish the two types of physics at the 95\% confidence level
based on the likelihood ratios.
We outline details of our procedure here,
and we present our results in section \ref{sec:results}.

At very low statistics in each bin,
as we expect to see if we restrict ourselves to high invariant mass events, we
maximize the maximum likelihood function
\begin{equation}
L(\mu_i,n_i) = \prod_i f(\mu_i,n_i),
\end{equation}
where $f(\mu,n) = e^{-\mu}\,\mu^n/n!$ is the standard Poisson distribution and the $\mu_i$ are the
means in each bin.  Using the 
standard $\chi^2$ distribution to perform fits is not valid,
as it assumes normally-distributed errors while the
correct underlying distribution is Poisson \cite{Baker:1983tu}.  
The mean in each bin is found by integrating the differential cross section over
the bin width. An unbinned analysis is also possible, but as the unbinned case can be obtained by
taking the limit of the bin width to zero, we  derive the results for a binned analysis and
comment further on appropriate bin sizes in Appendix \ref{sec:binning}.

The bin means are functionally dependent on the parameters of the theory; 
for instance, with contact interactions, we have $\mu_{i\,\mathrm c}=\mu_i(\Lambda)$, while for RS we
have $\mu_{i\,\mathrm r} = \mu_i(M_{\mathrm g},k/\overline M_{\mathrm{Pl}})$, with the $i$ index labeling bins.

With few high energy events, a better statistic than $\chi^2$ is \cite{Baker:1983tu}
\begin{equation}
Q = -2\ln\lambda=2\sum_i\left(\mu_i-n_i+n_i\ln\frac{n_i}{\mu_i}\right),
\end{equation}
where we define
\begin{equation}
\lambda = \frac{L(\mu_i,n_i)}{L(n_i,n_i)} < 1,
\end{equation}
which is the maximum likelihood ratio.  The statistic $Q$ has the property that $Q\in[0,\infty)$,
analogous to the $\chi^2$.
Maximizing $L$ is equivalent to minimizing $Q$, which is the criterion used 
for fitting, and good fits are characterized by small $Q$.  

There are some important differences however from
  $\chi^2$, which are seen when we apply the two statistics to a single bin (take $i$
to only have one value).  Unlike the $\chi^2$ distribution, the mean and
variance of $Q$ is dependent on the mean of this bin \cite{CDF:LL}.
For the $\chi^2$, we always have an expectation value $\langle\chi^2\rangle$/d.o.f.~(or
$\langle\chi^2\rangle$/bin)
of one, regardless of the bin mean.
With small bin mean, we instead find $\langle Q \rangle < 1$.  If we now sum over many bins,
the total $Q$ statistic will be dependent not only on the number of bins (like the $\chi^2$),
but also on the means of those bins.
Since the $Q$ distribution is dependent on the bin means, which in turn are functions of the parameters
of the theory, we have a different distribution of $Q$ for each set of parameters.  We therefore have
no choice but to
generate these through  Monte Carlo simulations.

We will assume in everything that follows that the correct underlying
new physics is a resonance model, and we will try to distinguish it from a contact
interaction. To do so, we compare the values 
of $Q$ that come from fitting resonance ($Q_{\mathrm r}$) and contact interaction ($Q_{\mathrm c}$)
models to \emph{any} simulated dataset. 
We compute
\begin{equation}
\Delta Q = Q_{\mathrm c} - Q_{\mathrm r} = 2\sum_i\left(\mu_{i\,\mathrm c}-\mu_{i\,\mathrm r}+n_i\ln\frac{\mu_{i\,\mathrm r}}{\mu_{i\,\mathrm c}}\right),
\end{equation}
expecting $\Delta Q<0$ for a dataset that looks more like a contact interaction and $\Delta Q>0$ for a
dataset that looks more like a resonance (the subscripts c and r stand for 
contact and resonance respectively). 

Statistical fluctuations can make a contact
interaction look more like a resonance and vice-versa, so it is not sufficient to examine the sign of
$\Delta Q$.  Our procedure for studying a resonance with mass $M$ and coupling $g$ is as follows:

\begin{enumerate}

\item Generate data $D$ for a pseudoexperiment with $1\,\,\mathrm{fb}^{-1}$ at
$\sqrt s = 7\,\,\mathrm{TeV}$ using a resonance model with parameters $M$ and $g$.

\item Perform fits of a resonance model ($M^D$ and $g^D$) and a contact interaction with
scale $\Lambda^D$ to the data
$D$, minimizing $Q^D_{\mathrm r}$ and $Q^D_{\mathrm c}$ respectively.  The superscript shows that
this is with respect to the data $D$.

\item Compute $\Delta Q^D = Q^D_{\mathrm c}-Q^D_{\mathrm r}$.

\item To determine the values of $\Delta Q$ that come from a contact interaction with scale
$\Lambda^D$ faking a resonance,
generate data $D'$ based on a contact interaction with scale $\Lambda^D$.
Fit resonance and contact
parameters to $D'$ and compute $\Delta Q^{D'}$.  Repeat many times to generate a probability
distribution $\mathcal P$ over possible values of $\Delta Q^{D'}$.  The corresopnding cumulative
distribution, given by the integral of $\mathcal P$, is $\mathcal C$.

\item To distinguish the resonance in the data $D$ from a contact interaction at
the 95\% confidence level, we require $\Delta Q^D > \mathcal{C}^{-1}(0.95)$.  In words, if
$\Delta Q^D$ is bigger than 95\% of the $\Delta Q^{D'}$ values based on fits to an underlying
conctact interaction, then we can say with confidence that we have a broad resonance
rather than a contact interaction.

\end{enumerate}

The maximum likelihood ratio is well-known and used frequently in particle physics experiments
in performing fits
and determining the parameters of a theory from data, but we are here using it to directly compare
two competing theories of new physics that may yield similar distributions of observables.


\subsection{Results of statistical analysis}\label{sec:results}

We now present the main results from our analysis.
We apply the maximum likelihood analysis from the last section
to the study of RS KK gravitons and non-universal
$B-3L_{\mu}$ $Z'$ discussed in section \ref{sec:models}.\footnote{In this section,
we do not consider the flavor-universal
$B-L$ case, as the LEP bounds imply narrow resonances in accessible regions of the parameter space
at the LHC.}  We determine over what regions of parameter space we can distinguish such models from contact
interactions.  Examining Fig.~\ref{fig:Contact}, it is apparent that the destructively-interfering
contact interactions will look most like a broad resonance.  Whether it is LL or VV that most resembles the resonance depends
on the nature of the resonance model's coupling.

The analysis
outlined in section \ref{sec:mlm} studies only a single $\Delta Q_{\mathrm{bf}}$ arising from a 
single simulation of a resonance in $1\,\,\mathrm{fb}^{-1}$ of data.  To quantify the effects
of the statistical fluctuations over many
different simulated experiments, we define {\bf reliability}
as the percentage of experiments in which we can distinguish a resonance from a contact interaction
at the 95\% confidence level.  This essentially tells us the probability of being able to
distinguish a resonance with $1\,\,\mathrm{fb}^{-1}$ at the LHC for various possible parameters.

\begin{figure}[t]
\begin{center}
\includegraphics[scale=0.9]{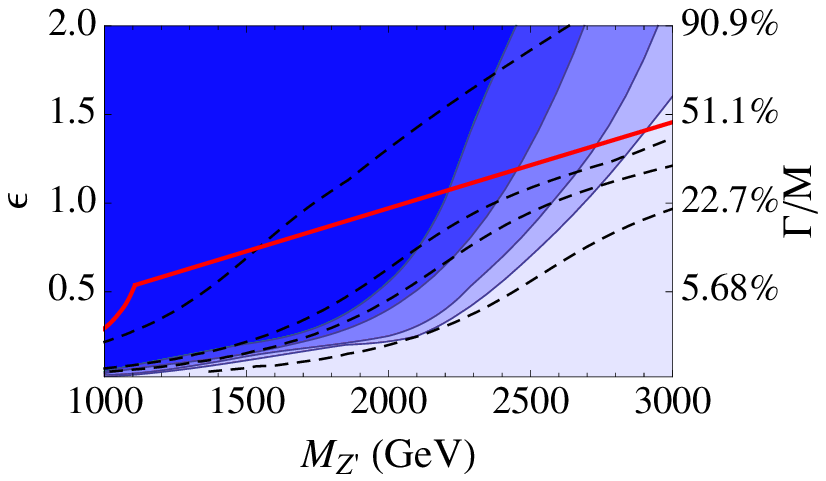}
\includegraphics[scale=0.9]{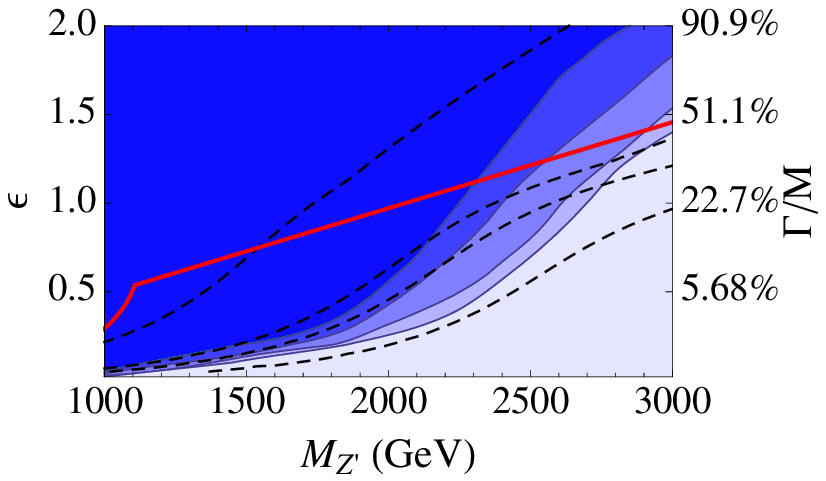}\\
\includegraphics[scale=0.9]{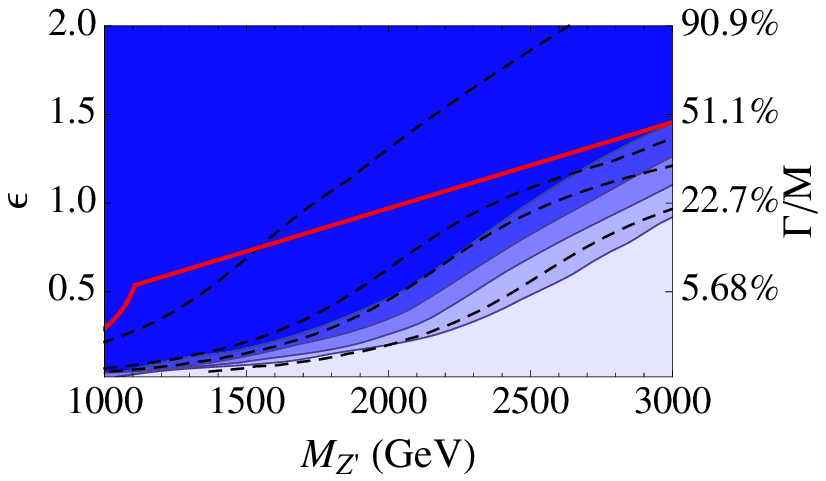}
\includegraphics[scale=0.9]{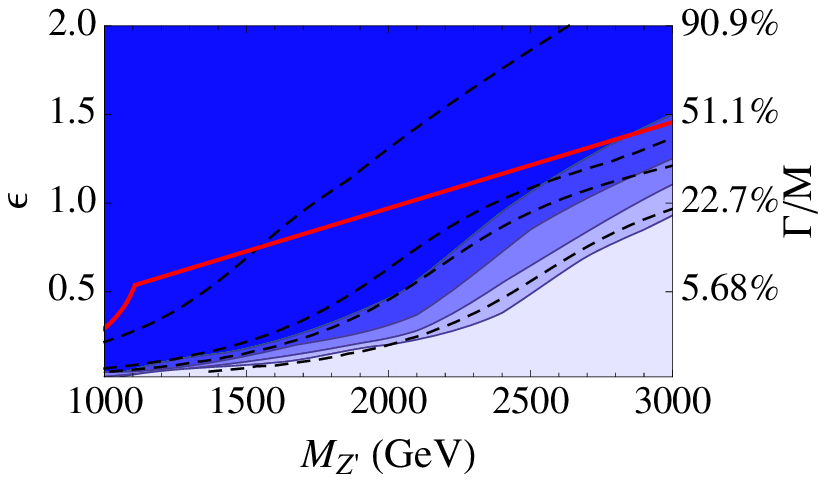}
\caption{Reliability of distinguishing $B-3L_{\mu}$ $Z'$ resonance at 95\% confidence level
from VV destructive (top left), LL destructive (top right), VV constructive (bottom left), 
and LL constructive contact interactions.  The shaded regions, from darkest to lightest, 
show regions with reliability: $>$ 99\%, 90-99\%, 70-90\%, 50-70\%, $<$ 50\%.  Excluded
regions are to the left of the solid
red line, while the dashed lines show cross
sections: 100 fb (top), 10 fb, 5 fb, 1 fb (bottom).}
\label{fig:ZPstat}
\end{center}
\end{figure}
\begin{figure}[t]
\begin{center}
\includegraphics[scale=0.9]{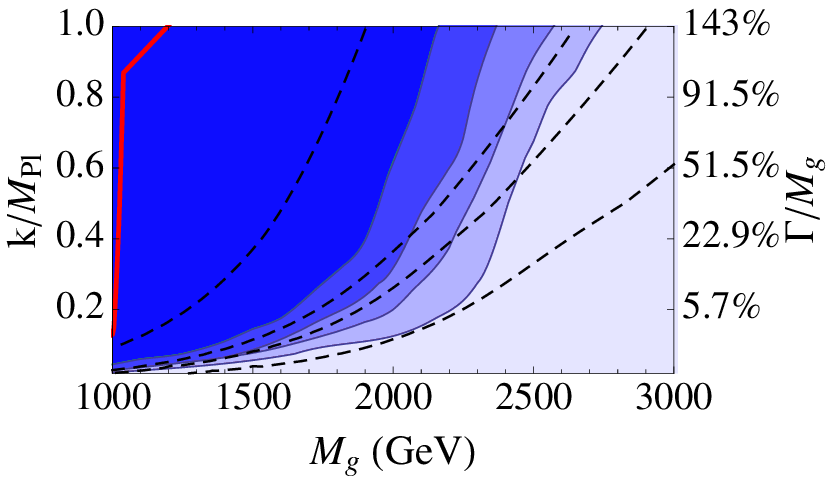}
\includegraphics[scale=0.9]{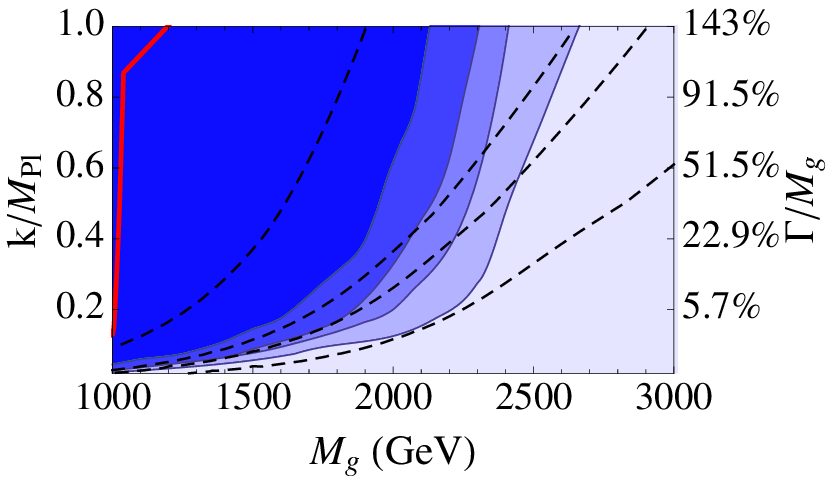}\\
\includegraphics[scale=0.9]{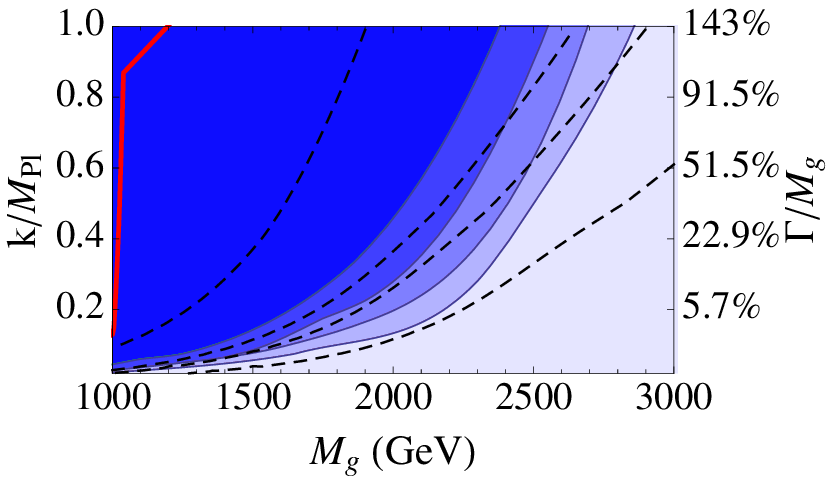}
\includegraphics[scale=0.9]{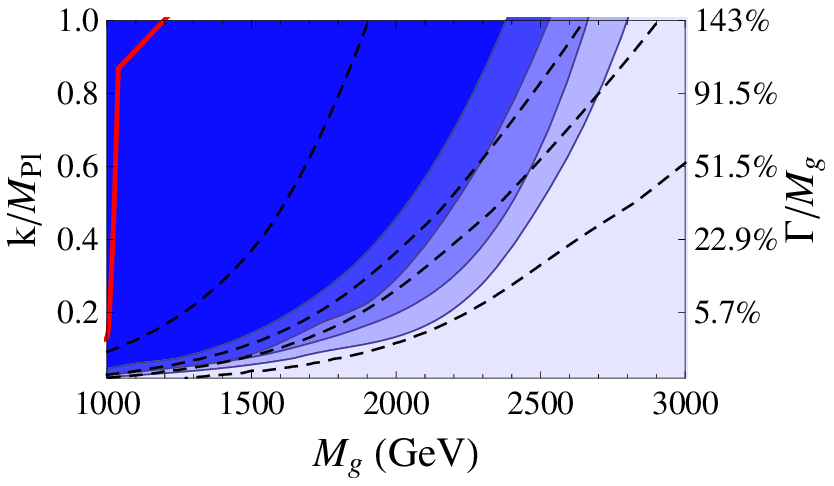}
\caption{Reliability of distinguishing RS resonance at 95\% confidence level from VV destructive 
(top left), LL destructive (top right), VV constructive (bottom left), and LL constructive contact 
interactions.  The shaded regions, from darkest to lightest, show regions with reliability: 
$>$ 99\%, 90-99\%, 70-90\%, 50-70\%, $<$ 50\%. Excluded
regions are to the left of the solid
red line, while the dashed lines show cross
sections: 100 fb (top), 10 fb, 5 fb, 1 fb (bottom). }
\label{fig:RSstat}
\end{center}
\end{figure}

Our results, showing the reliability curves for distinguishing resonances from contact interactions,
are presented in Fig.~\ref{fig:RSstat} for RS and in Fig.~\ref{fig:ZPstat} for the non-universal
$Z'$. Some of the key points from the figures that are true for both models are:
\begin{enumerate}

\item It is most difficult to distinguish a resonance from a contact interaction that interferes
destructively with the SM (in agreement with our earlier reasoning).  For $Z'$, it is the VV
destructive contact interaction that is most likely to fake the resonance since that is the nature
of the $Z'$'s own interference and coupling.  For RS, where SM interference does not modify the
total cross section, both destructive models do equally well,
with only modest differences between
the others.

\item For narrow resonances ($k/\overline M_{\mathrm{Pl}} < 0.4$, $\epsilon<0.75$),
the reliability curves track the cross section contours.  Here, the
relevant question is only the number of signal events required for detection:
with five or more events,
we can reliably distinguish between a resonance and contact interaction.

\item  For broad, \emph{low-mass} resonances 
($k/\overline M_{\mathrm{Pl}} > 0.4$ and $M < 2$ TeV for RS,
$\epsilon > 0.75$ and $M < 2.5$ TeV for $Z'$),
we can universally discriminate between contact
interactions and resonances because of the large number of events at
$1\,\,\mathrm{fb}^{-1}$. The effects of statistical fluctuations become negligible with
many events.

\item   As expected, broad, \emph{high-mass} resonances
($k/\overline M_{\mathrm{Pl}} > 0.4$ and $M > 2$ TeV for RS,
$\epsilon > 0.75$ and $M > 2.5$ TeV for $Z'$)
are the hardest to 
distinguish from contact interactions at lower statistics.
We find that it is still possible to detect broad 
resonances with confidence over most of the kinematically accessible parameter space, but the reliability of doing
so diminishes when the width is greater than $\Gamma/M\sim 20\%$. 
\end{enumerate}

\begin{figure}[t]
\begin{center}
\includegraphics[scale=0.9]{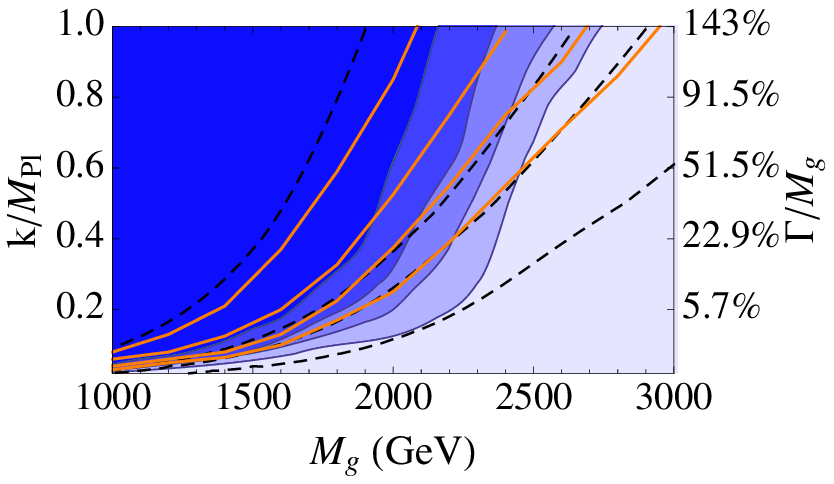}
\includegraphics[scale=0.9]{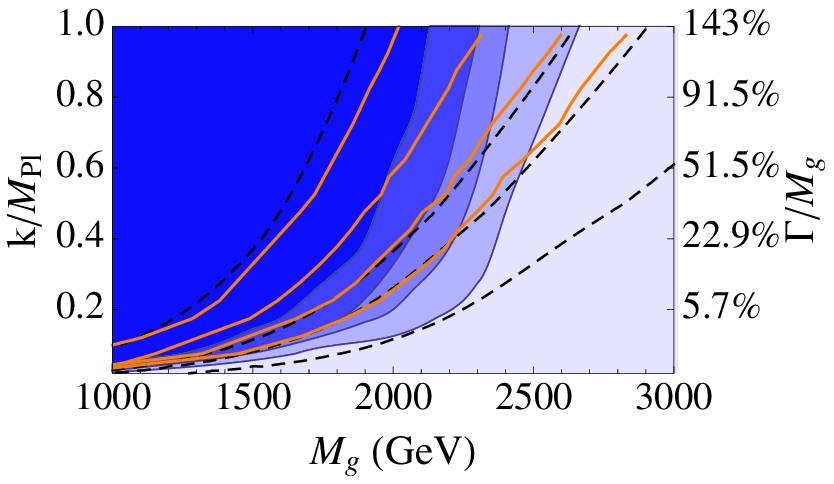}
\caption{Plots showing mean best-fit values of $\Lambda$ for RS models compared to VV (left) and LL
(right) destructive contact interactions.  Contours of $\Lambda$ are shown in solid
orange lines.  They are for the left plot: 7 (top),
8.5, 10, 11.5 TeV (bottom); for the right plot:
5 (top), 6, 7, 8 TeV (bottom).  The reliability regions are shaded
and cross sections shown with dashed lines, as in Fig.~\ref{fig:RSstat}.}
\label{fig:Lambda}
\end{center}
\end{figure}

The analysis described above was done only for a single, isolated resonance.  Generally, when we
encounter strongly-coupled physics, we may find multiple resonances.  We certainly
expect to see additional resonances in the RS framework, namely the higher KK modes.
The interference between resonances may be sufficiently high that
we can no longer neglect the heavier modes, depending on resonance width and spacing.

As a check, we compute the contribution of multiple graviton
KK modes (and their mutual interference) to the Drell-Yan process in RS.  For 
$k/\overline M_{\mathrm{Pl}}<0.5$, the effects of higher KK modes on the differential and total
cross sections around $M_{\mathrm g}$ are minimal.  At higher coupling, we see enhancements to the
cross section of up to $\sim 15$ \%; most importantly, however, \emph{the shape is not distorted}.
Thus, these effects do not substantially change the outcome of the single-resonance analysis;
the marginal increase
in event rate will make at best make model discrimination slightly easier.

To better understand how the fitting procedure works, we plot the mean best-fit values of $\Lambda$
for RS resonances in Fig.~\ref{fig:Lambda}, both for LL and VV destructive contact interactions.
It is apparent that the best-fit values of $\Lambda$ are strongly correlated with the
total number of events.  The contact interaction most likely to fake a broad resonance is one
that predicts the \emph{same} number of events as the resonance, only with the events
distributed in a different way.
This is in accordance with our picture of unfortuitously distributed upward- and
downward-fluctuations making a contact interaction look like a resonance.   

In our analysis, we choose a bin size of 100 GeV, since that is on the order of the detector resolution
for dimuon invariant masses around 1 TeV \cite{Kortner:2007qj}.  The range of invariant masses that we
consider is 300 GeV to 4 TeV.\footnote{Since the area around the $Z$ peak is used to normalize the cross
section and match the PDFs, we choose a lower value that safely avoids this region}  We have not included
detector smearing effects.  A discussion of the effects of including detector smearing and varying
the bin size (justifying a binned analysis) is found in the Appendices.


\subsection{Electrons and photons}
The non-universal $Z'$ we have been considering evades the LEP bounds by coupling only to muons,
and there will be no excess dielectron or diphoton final state events over the SM prediction.
RS gravitons, however, couple to both electrons and photons, and these should make it easier
to distinguish gravitons from contact interactions.

Electron and photon final states, like muons, have small SM backgrounds and are straightforward
to detect.  Depending on cuts imposed to minimize fake rates, detector efficiencies for electrons
and photons can be as high as 60-85\% \cite{ATLAS:expected}.  The cross section for photon production by a KK graviton
is twice that for muons, and so the total event rate increases by just under a factor of four
when we include all channels (it is smaller than four because of the lower efficiencies for
photons and electrons).
\begin{figure}[t]
\begin{center}
\includegraphics[scale=1]{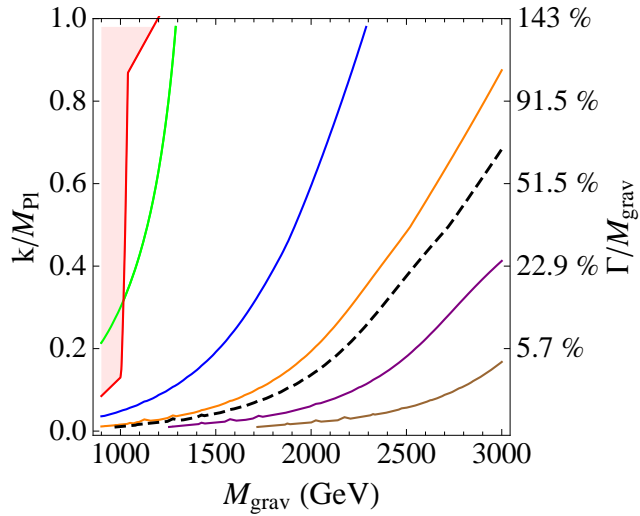}
\caption{Plots showing dimuon cross section for the first KK mode of the graviton in
RS models (to muons, electrons, and photons) 
at $\sqrt s=7\,\,\mathrm{TeV}$.
Legend: green is 1 pb, blue is 100 fb, orange is 10 fb, purple is 1 fb.
The black curve indicates the cross section for 5 events. }
\label{fig:RSxsecEW}
\end{center}
\end{figure}

We plot our results for the RS cross section to electrons, muons, and photons in
Fig.~\ref{fig:RSxsecEW}.  We do not repeat the full statistical analysis and
comparison to contact interactions because a) we expect the qualitative result to
be the same, and b) it is difficult to systematically explore the full range of
possible contact interactions, since we could have different coefficients for the 
operators involving different final states, etc.  However, we do compare the cross
section to that for muons alone and find that the new cross section contours are
now shifted upward in mass by about 400 GeV.  We therefore expect to be able to 
probe resonances to 400 GeV higher masses than if we relied only on muons, and
the reliability regions would also shift accordingly, meaning that we have
difficulty discriminating 2.8 TeV gravitons from contact interactions, rather
than the 2.4 TeV gravitons we had with just muons.

\section{Angular distribution}
In this section we demonstrate how to use the 
pseudorapidity ($\eta$) distribution of the final state high $P_T$ muons to provide
another handle in  discriminating between models.  
We define a new variable called ellipticity, which measures how
longitudinal the outgoing muons are in either the forward or backward directions, 
with no need to explicitly identify the direction of the quark vs. antiquark 
(as would be necessary for forward-backward asymmetry determination). 
We will see that this is a more efficient way to identify
parity violation in new physics, particularly for broad resonances. 
 
\begin{figure}[t]
\begin{center}
\includegraphics[scale=0.5]{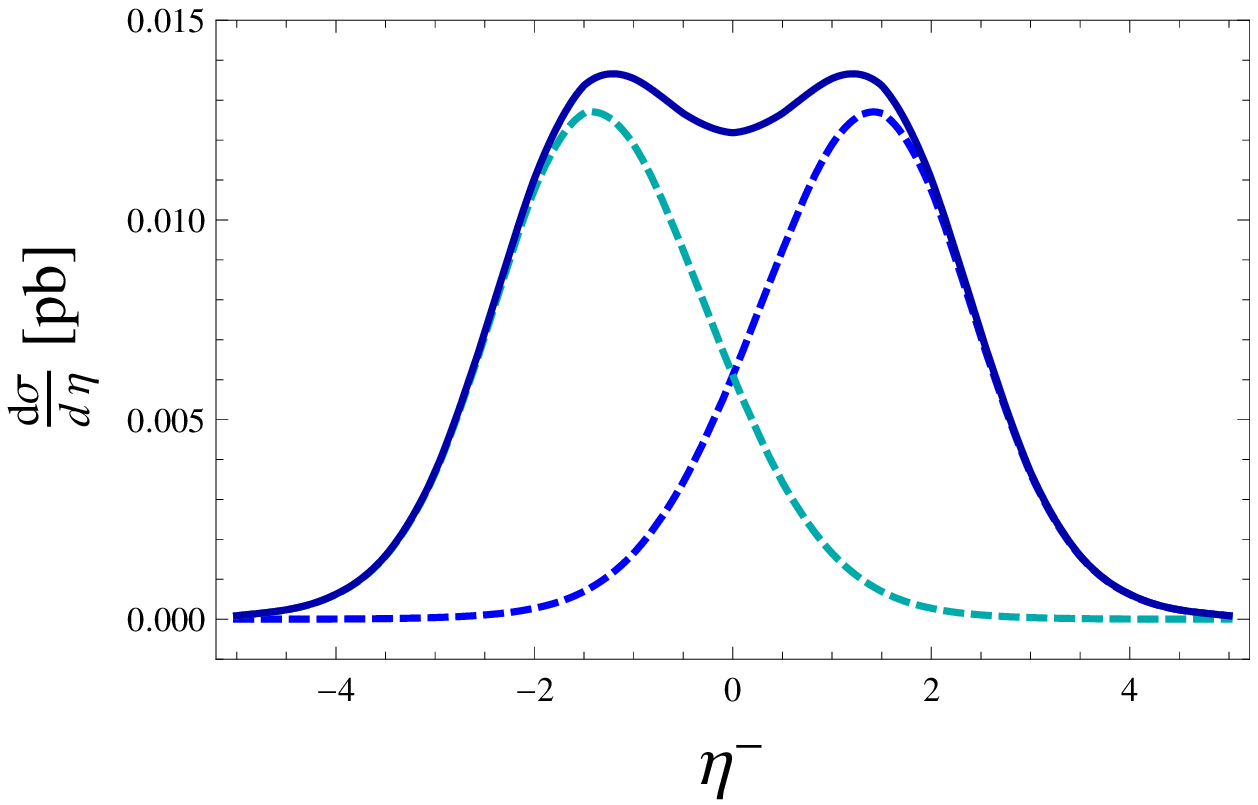} \hspace{1cm}
\includegraphics[scale=0.5]{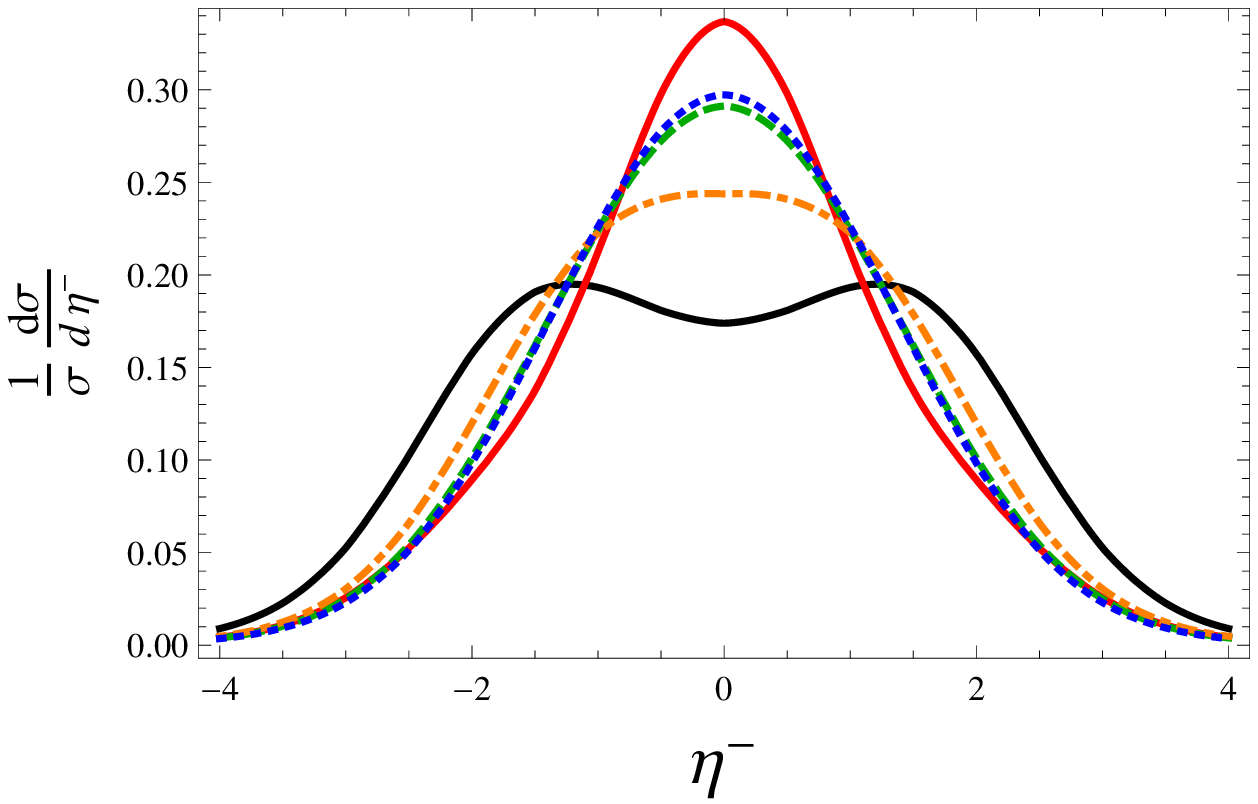}  \\ \vspace{2mm}
\includegraphics[scale=0.5]{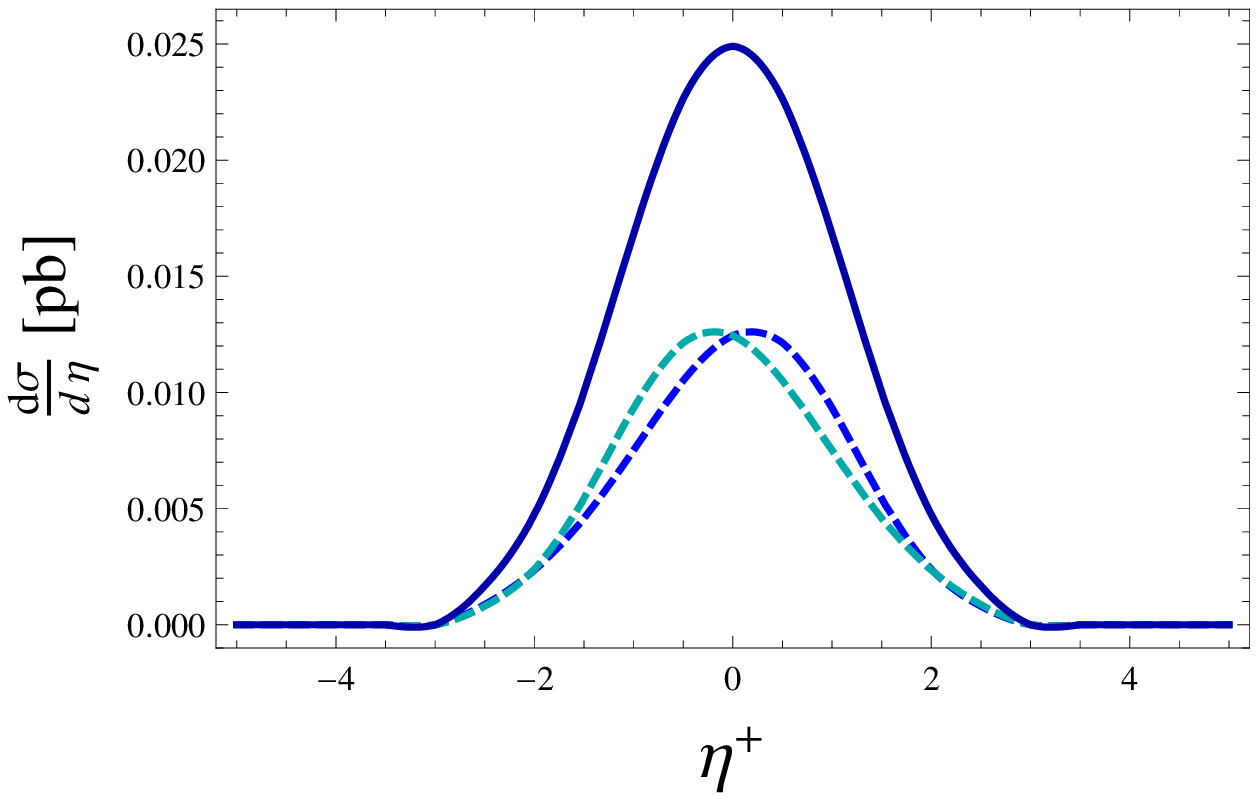} \hspace{1cm}
\includegraphics[scale=0.5]{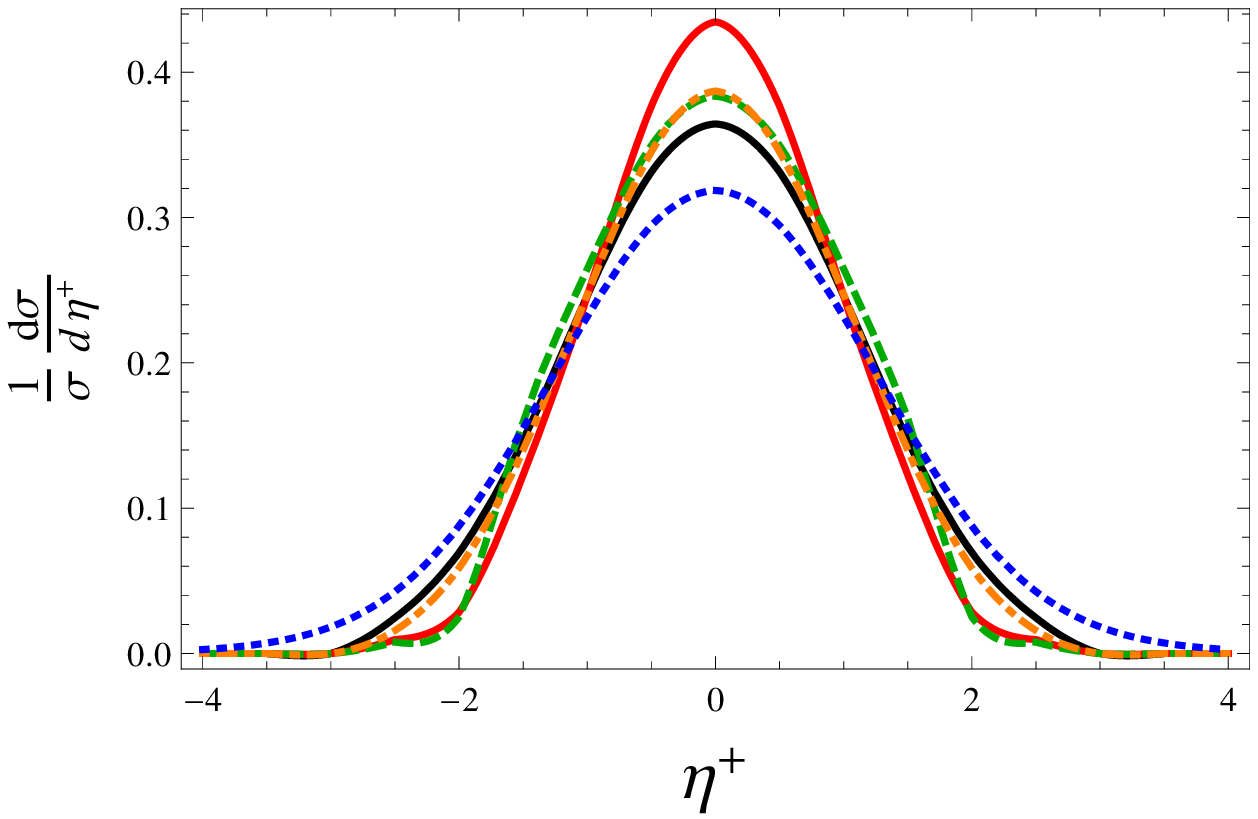} 

\caption{
    The left plot shows the hadronic differential cross section in $\eta^-$ (upper) and $\eta^+$ (lower)
    for 
    $pp \to \mu^+ \mu^-$ via the standard model at $E_{\rm cm} = 7$ TeV with $M_{\mu \mu} > 400$ GeV.
    The dashed curves represent the individual contribution from each of the two terms in 
    Eq.~\ref{diff:xsec} and the solid curve is the sum. 
    The right plots shows the same cross sections 
    for the SM (solid black),
    RS model with $\Mg = 1300$ GeV, $k/\MPl = 0.3$ (solid, red), 
    ($B-3L_\mu$) $Z'$ model with $\Mg = 1300$ GeV, $k/\MPl = 0.48$ (dashed, green), 
    LL composite model with destructive interference and $\Lambda = 4550$ GeV (dotted,blue),
    and
    VV composite model with destructive interference and $\Lambda = 6150$ 
    GeV (dot-dashed, orange).
    They have been normalized to compare shape.
}
\label{EtaPlot}
\end{center}
\end{figure}

The SM violates parity, whereas new physics models may not.  
In a $pp$ collider, this is encoded in the shape of the pseudorapidity distribution, 
as can be seen in Fig.~\ref{EtaPlot}, which shows the hadronic differential cross sections 
($d\sigma/d \eta^{\pm}$) for the standard model, where we have defined $\eta^{\pm}$ as the 
pseudorapidity of the outgoing $\mu^{\pm}$ from $pp \to \mu^+ \mu^-$.
The outgoing $\mu^-$s are preferentially scattered forward relative to the quark in the SM,
and since the quark could come from either proton, 
the $\eta^-$ distribution for the SM demonstrates a characteristic double peak structure.
On resonance, new physics contributions that respects parity, such as the RS model, 
would wash out this effect, resulting in a transversely peaked distribution as shown on 
the right side of Fig.~\ref{EtaPlot}.  Note that, at a $pp$ collider, studying the 
forward backward asymmetry in the CM frame will not capture this effect since, 
without identifying the quark direction, the symmetry of the $\eta^{\pm}$ distribution
would make the distinguishing peak structure cancel.

\begin{figure}[t]
\begin{center}
\includegraphics[scale=0.50]{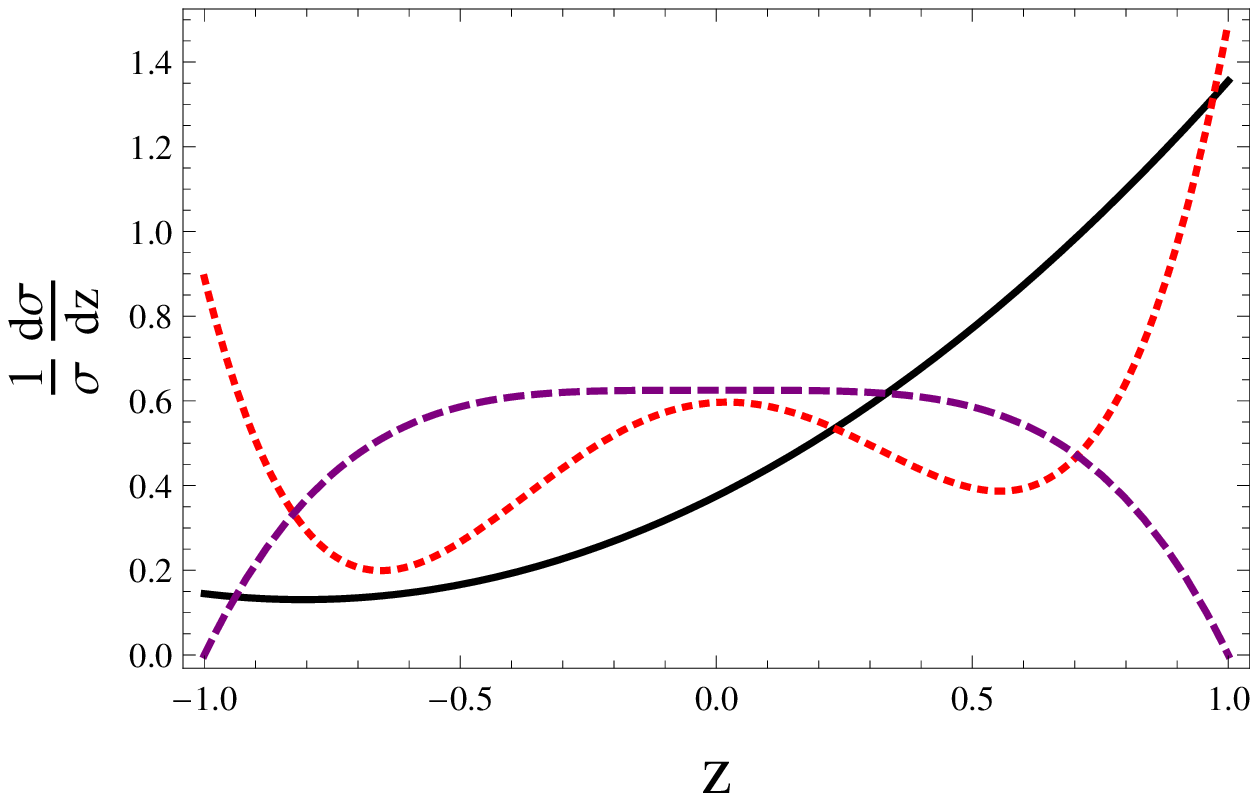} 
\includegraphics[scale=0.50]{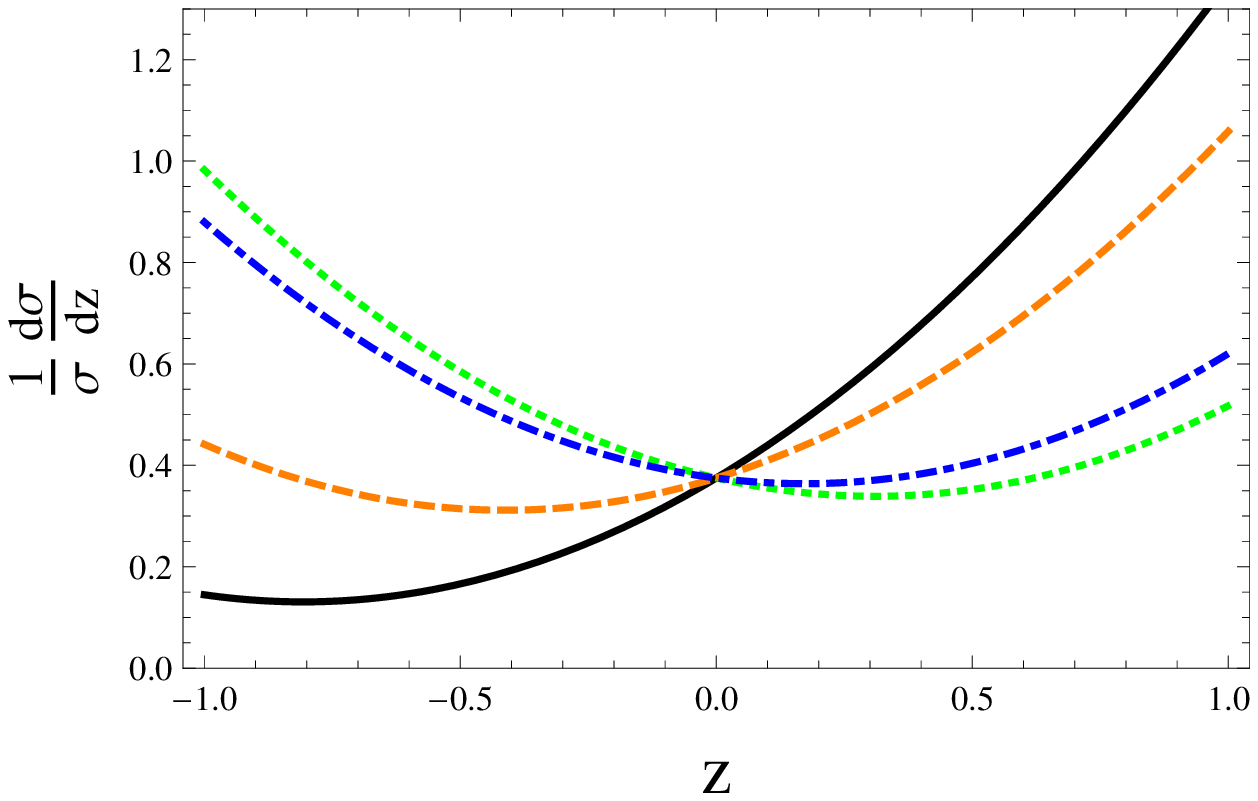} 
\caption{
The normalized differential partonic differential cross section for $u\bar{u} \to \mu^+ \mu^-$ via 
the standard model is shown in both plots (solid black).
The left plot shows the cross section for the SM with RS graviton 
of mass $\Mg = 1300$ GeV and coupling $k/\MPl = 0.3$ for the processes
$u\bar{u} \to \mu^+ \mu^-$ (dotted red),
and $gg \to G^{\ast} \to \mu^+ \mu^-$ (dot-dashed purple).  
The right plot shows the differential cross section for the process $u\bar{u} \to \mu^+ \mu^-$
for the SM plus ($B-3L_{\mu}$) $Z'$ with mass $\Mzp = 1300$ GeV and coupling $\e = 0.48$ (dotted green),
SM plus LL (destructive) composite model with $\Lambda = 4550$ GeV (dashed orange),
and SM plus VV (destructive) composite model with $\Lambda = 6150$ GeV (dashed orange).
All plots are shown with partonic CM energy $\sqrt{\hat{s}} = 1100$ GeV.
}
\label{fig:Theta}
\end{center}
\end{figure}

The shape of the $\eta^-$ distribution is significantly different 
than that of the $\eta^+$ distribution and requires explanation.
The underlying hard interaction involves 
a valence quark whose momentum fraction is generally much larger than the antiquark's 
and the resulting collision is boosted in the quark's direction. Furthermore,   
the Standard Model preferentially scatters $\mu^-$s into the same direction as the quarks as shown
in Fig.~\ref{fig:Theta}, and when combined with the effect of the large boost, 
the resulting $\eta^-$ distribution is peaked away from the zero as shown by the dashed lines 
in Fig.~\ref{EtaPlot}.  
Since the valence quark could come from either proton, the distribution must be symmetric in  
$\eta^-$ and the two dashed curves are added to give the final double arched curve in Fig.~\ref{EtaPlot}.  
The $\mu^+$ cross section is more centrally
peaked since the $\mu^+$ is preferentially scattered backwards from the quark and, after the boost, 
becomes more transverse with smaller rapidity.  
A $p\bar{p}$ collider would separately show the effects of parity violation in both of the $\eta$ distributions 
since these curves would are symmetric; 
however, the large boost at a $pp$ collider magnifies the effect in the $\eta^-$ distribution 
while simultaneously diminishing the effect in $\eta^+$.

Therefore, we study a new observable, the {\bf ellipticity} ($E_\eta$), 
to probe the shape of the $\eta^{-}$ distributions and we demonstrate its usefulness in 
discriminating between new physics
models at the LHC.  
We calculate the  ellipticity with the  formula
\begin{equation}
\label{Eeta_def}
E_{\eta} = 
     \frac{\ds 
        \left[ 
            \int_{-x}^{x} 
            - \left( 
                  \int_{-\eta_{\rm max}}^{-x} 
                + \int_{x}^{\eta_{\rm max}} 
              \right) 
        \right]
        d\eta^-  \dfrac{d\sigma}{d \eta^-} }
        {\ds \int_{-\eta_{\rm max}}^{\eta_{\rm max}}   d\eta^-  \dfrac{d\sigma}{d \eta^-} },
\end{equation}
where $\eta^{\pm}$ is the pseudorapidity of the outgoing $\mu^{\pm}$.
This quantity seeks to exploit the distinctive shape of the $\eta^-$ distributions.
The end result is relatively insensitive to 
values of $x \sim 1$ and $\eta_{\rm max} \sim 2.5$, and so we have chosen
$x = 1.0$, $\eta_{\rm max} = 2.5$ for this study. Ultimately, the value of $x$ could be optimized.
Furthermore, studying a ratio has the usual advantage that some systematic uncertainties,
such as the K-factor, integrated luminosity, PDF uncertainties, etc., will be common to 
both quantities and thus divide out.

We will also contrast ellipticity with the {\bf center-edge asymmetry} 
\cite{Osland:2008sy} ($A_{\rm ce}$),
a related quantity based on the angular distribution in the CM frame that
has been considered for spin determination.
The center-edge asymmetry is defined using the quantity 
$z = \cos \theta^{\ast} = \tanh(\frac{\eta^- - \eta^+}{2})$ and is 
\begin{equation}
\label{Ace_def}
    A_{ce} = 
    \frac
    { \ds 
        \left[ 
            \int_{-z^{\ast} }^{z^{\ast} } 
            - \left(  \int_{-z_{\rm max} }^{-z^{\ast} } 
            +  \int_{ z^{\ast} }^{z_{\rm max}} \right)  
        \right] 
        dz \frac{d \sigma}{dz} }
    { \ds \int_{-z_{\rm max}}^{z_{\rm max}} dz \frac{d \sigma}{dz} }.
\end{equation}
This quantity attempts to exploit the differences in the $\cos \theta^{\ast}$ 
distribution due the spin of the structure of the signal.  
For our study, we have taken $z_{\rm max} = 1.0$ and $z^{\ast} = 1/2$. 
A more thorough study of the variation of $z^{\ast}$ was carried out in 
\cite{Osland:2008sy}.  
Also, a modified version of center-edge asymmetry ($\tilde{A}_{\rm ce}$) was proposed
in \cite{Diener:2009ee} that is based on $\Delta \eta = \eta^- - \eta^+$ instead of $\cos \theta^{\ast}$.  
Since $\cos \theta^{\ast} = \tanh (\Delta \eta/2)$, $A_{\rm ce}$ and $\tilde{A}_{\rm ce}$ behave 
similarly and any conclusions we draw for $A_{\rm ce}$ will also apply to $\tilde{A}_{\rm ce}$.
Notice that while $A_{\rm ce}$ and $E_{\eta}$ are defined analogously,
with $z \leftrightarrow \eta^-$.  We will demonstrate that, while 
$E_{\eta}$ and $A_{\rm ce}$ are both sensitive to the underlying spin structure,
$E_{\eta}$ provides more discriminating power among models with 
different parity violating character.

\begin{figure}[t]
\begin{center}
\includegraphics[scale=0.50]{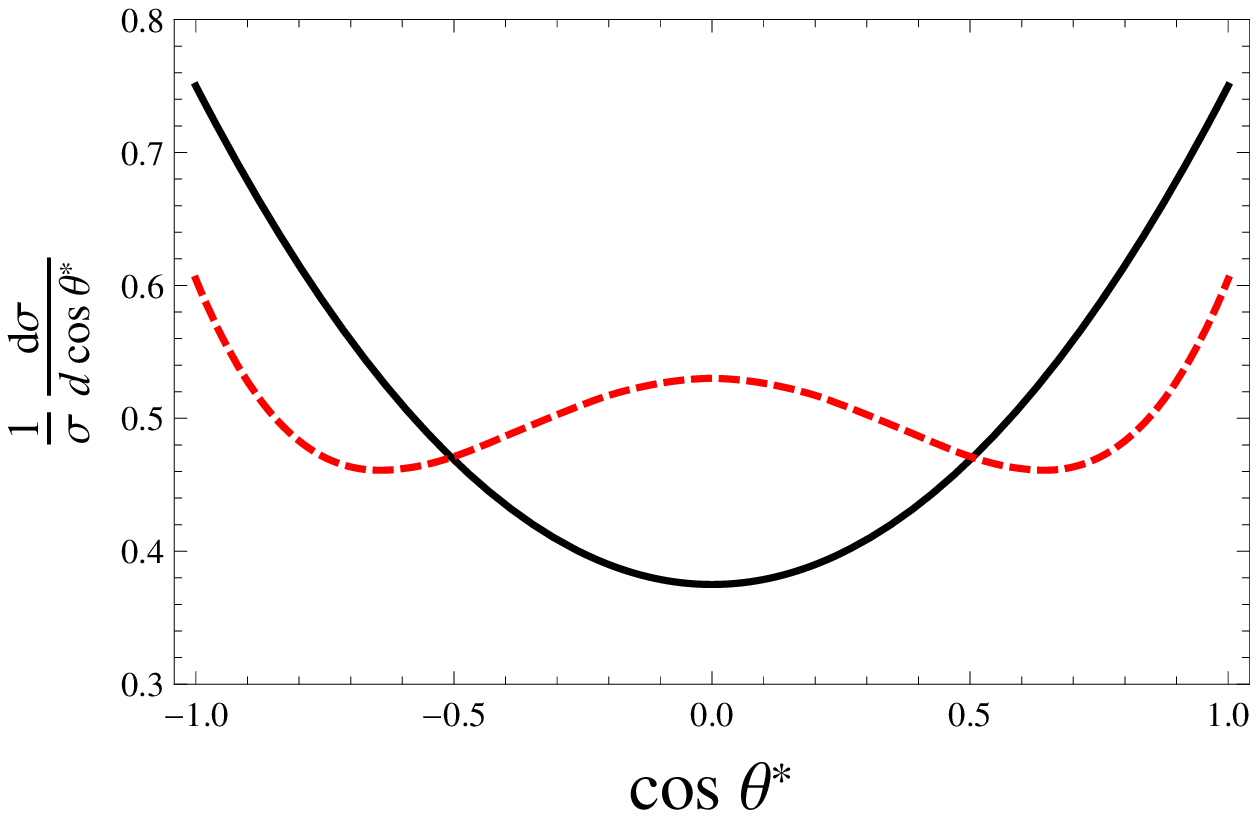} 
\includegraphics[scale=0.50]{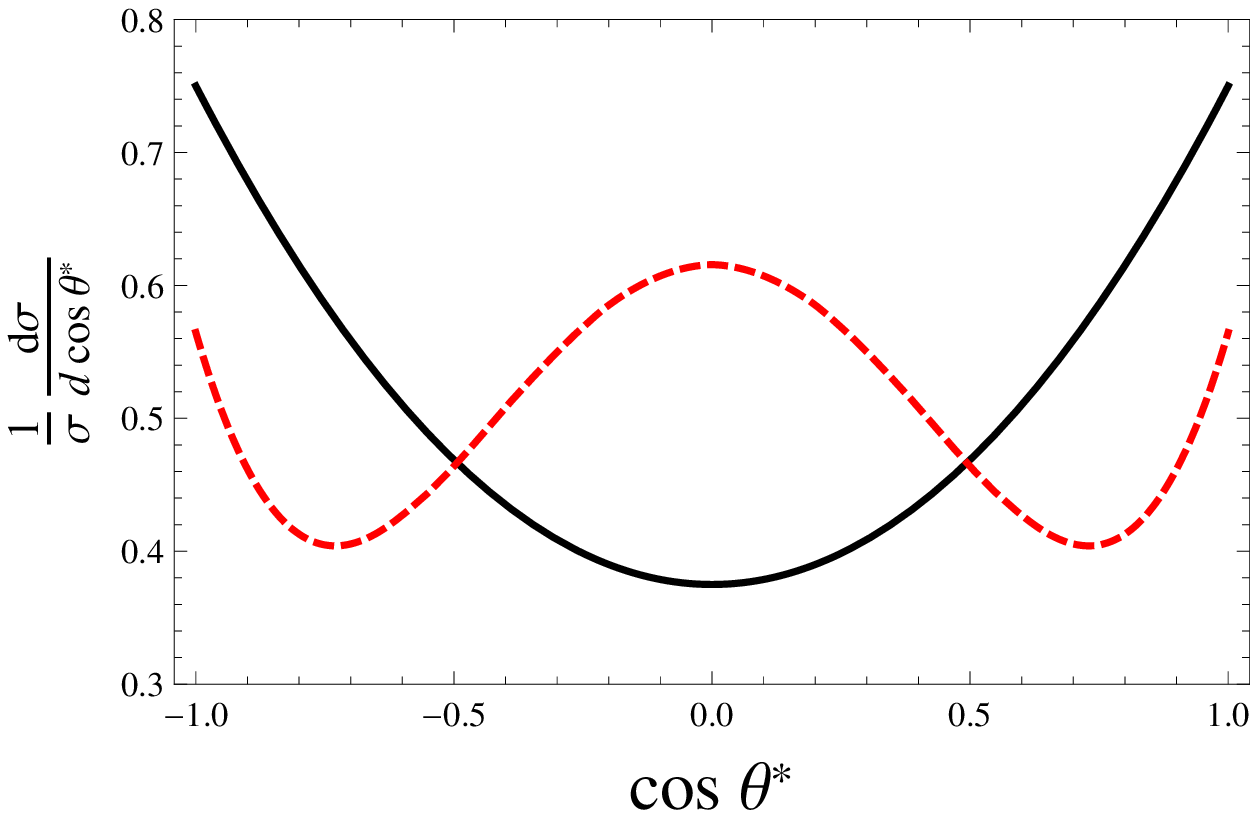} 
\caption{. 
The (hadronic) differential cross section in $\cos \theta^{\ast}$ for $pp \to \mu^-\mu^+$ is shown 
at $E_{\rm cm} = 7$ TeV with $M_{\mu\mu} > 400$ (left) for the same models as Fig.~\ref{fig:Theta}.
The right plots shows the same cross section for the same models as Fig.~\ref{E:3} 
at $E_{\rm cm} = 10$ TeV with $M_{\mu\mu} > 1000$. 
The RS model corresponds to the dashed red curve and shows a distinctive shape due to the spin-2 character 
of the graviton and its coupling to
initial state gluons.  
The other models are shown in solid black.
}
\label{fig:hTheta}
\end{center}
\end{figure}

Spin determination studies of narrow resonances have focused on the distribution of
$\cos \theta^{\ast} = \tanh (\eta^- - \eta^+)/2$,
where $\theta^{\ast}$ is the center of mass frame partonic scattering angle between the quark and the $\mu^-$.
Quantities that rely on $\cos \theta^{\ast}$ 
differ from quantities that rely on $\eta^{-}$ for several reasons.

\begin{itemize}
\item
$\cos \theta^{\ast}$ depends on a difference of rapidities, which makes the quantity boost invariant.
The boost magnifies
the effects of parity violation and, since $\cos \theta^{\ast}$ is boost invariant, 
the characteristic double peak feature of the standard model 
$\eta^-$ distribution is not present for $\cos \theta^{\ast}$.  

\item 
The forward-backward asymmetry ($A_{FB}$) depends on $\cos \theta^{\ast}$ and
requires that one determine the direction 
from which the quark originated.  This direction is not known in a $pp$ collider
{\it a priori} and the assignment of $\theta^{\ast}$ for each event is ambiguous.  
For very forward events, the boost required to bring the muons to the CM frame aligns with
the direction of the quark's momentum \cite{Dittmar:1996my}. 
A rapidity cut to remove the low rapidity events would
generate a sample for which we could measure $A_{FB}$; however, it would also remove a majority 
of the signal events, which tend to be centrally peaked, and thus would not be useful 
until a significant amount of data has been recorded.

Another version of $A_{FB}$ was proposed in \cite{Diener:2009ee} which probes differences in the distribution 
of $|\eta^-| - |\eta^+|$. 
Since this variable is sensitive to the boost of the system, it behaves similarly to $E_{\eta}$. 
Moreover, they showed that the inclusion of transverse events did not significantly 
affect the discriminating power of this variable and that it does not suffer
the deficiency of the traditionally defined version of $A_{FB}$ based on $\cos \theta^{\ast}$.

\item
$A_{\rm ce}$ cannot distinguish between models with the same spin structure, but with different amounts 
of parity violation.  For example, the composite models being considered arise from integrating out 
a heavy spin one resonance, and so they share the same initial state and spin structure to the $Z'$ models.
After convolving the partonic differential cross sections with the parton luminosity
function, any parity violating effects are washed out as demonstrated 
in Fig.~\ref{fig:hTheta}. These models are indistinguishable using $A_{\rm ce}$.  
Conversely, not only can $E_{\eta}$ can distinguish models with similar spin structure, but with different 
amounts of parity violation.

\item
A difference in 
$A_{\rm ce}$ indicates either a different spin structure and/or coupling to initial state gluons.
Thus $A_{\rm ce}$ can identify resonances mediated 
by an RS graviton, but it cannot distinguish between the rest of the models that we considered.
In this regard, $A_{\rm ce}$ and $E_{\eta}$ are similar.

\begin{figure}[t]
\begin{center}
\includegraphics[scale=0.4]{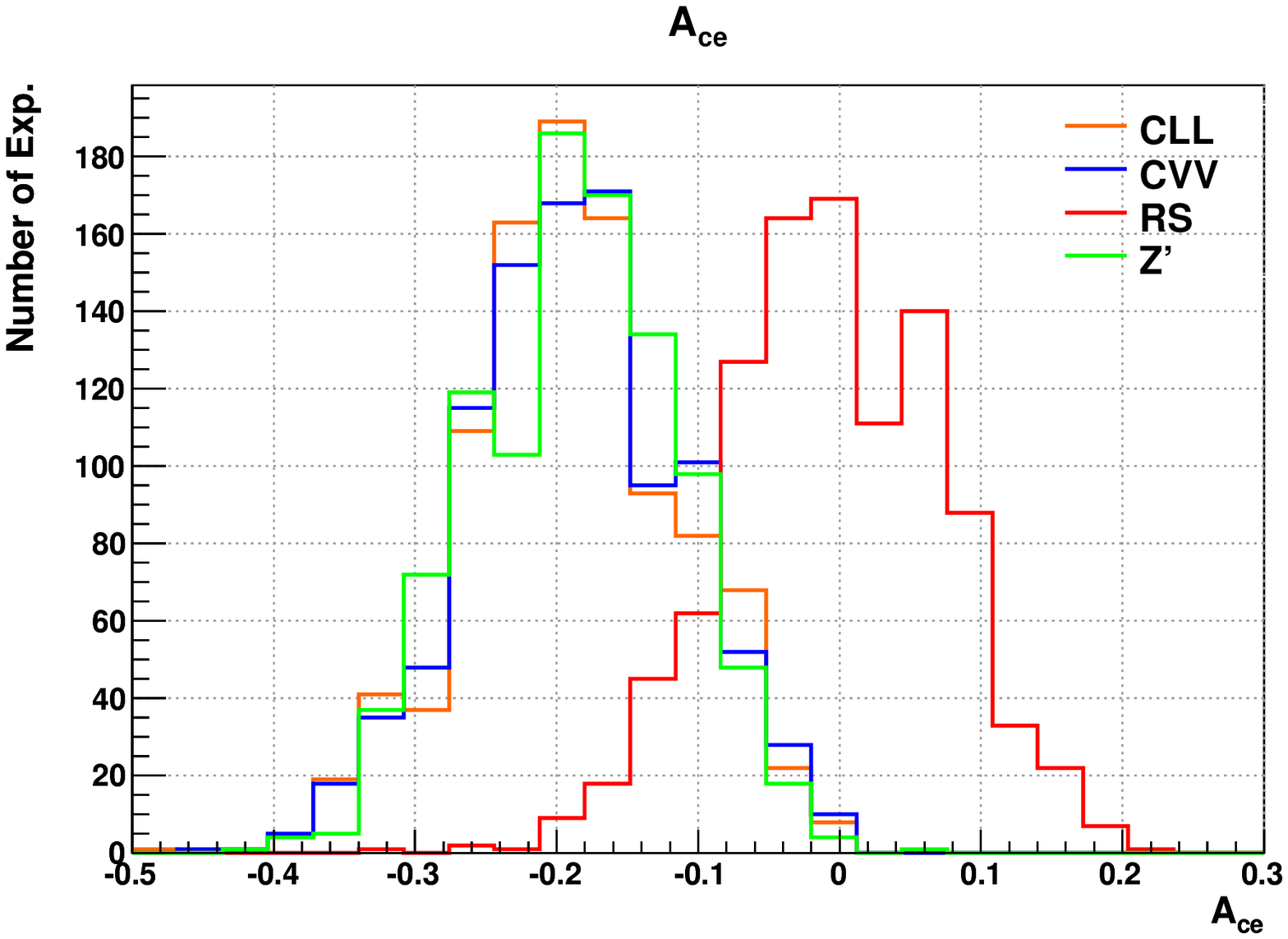}
\includegraphics[scale=0.4]{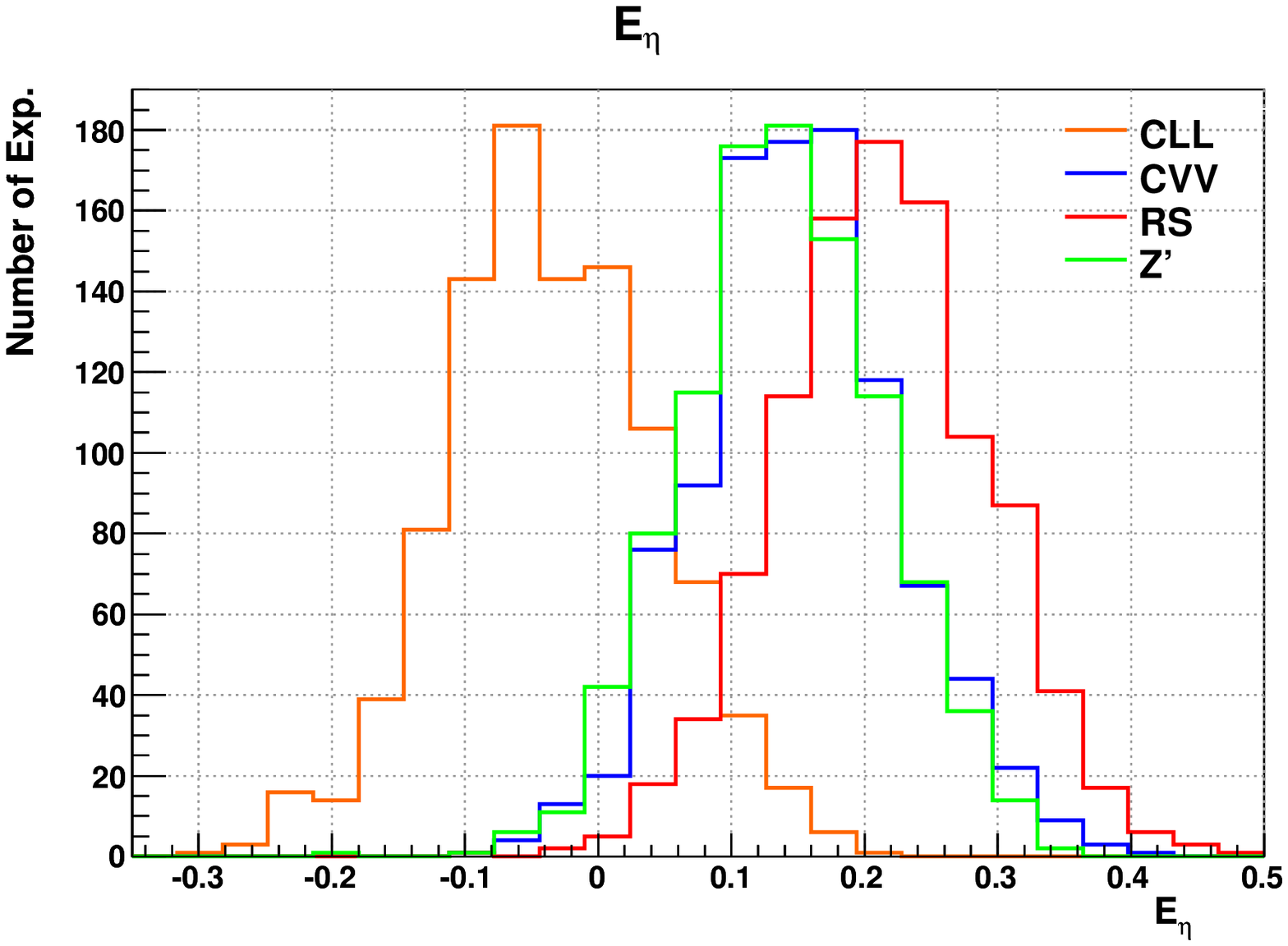}
\caption{$A_{\rm ce}$ (left) and $E_{\eta}$ are shown for 1000 pseudo-experiments for  
    $E_{\rm cm} = 7$ TeV, $\int dt \mathcal{L} = 1$ fb$^{-1}$  with $M_{\mu \mu} > 400$ GeV.
    The models shown are a
    RS model with $\Mg = 1300$ GeV, $k/\MPl = 0.3$ (red), 
    ($B-3L_\mu$) $Z'$ model with $\Mg = 1300$ GeV, $\e = 0.48$ (green), 
    LL composite model with destructive interference and $\Lambda = 4550$ GeV (blue),
    and
    VV composite model with destructive interference and $\Lambda = 6150$ GeV (orange), 
    }
\label{E:1}
\end{center}
\end{figure}

\end{itemize}

The calculation of the ellipticity is straightforward.
The following expression gives the differential cross section in $M_{\mu \mu}$ and $y^{\pm}$,
where $y^{\pm}$ is the rapidity of $\mu^{\pm}$:
\begin{align}
\label{diff:xsec}
\frac{d\sigma}{dM^2 dy^{-} dy^{+} } 
= 
    \frac{1}{E_{\rm cm}^2}   
     \frac{1}{2\cosh^2 y^\ast} 
     \frac{1}{1+\delta_{ij}} 
     &
     \left\{
       f_{i}(\sqrt{\tau} e^{-Y} )f_{j}( \sqrt{\tau} e^{Y} ) 
        \frac{ d \hat{\sigma}_{ij}}{d \cos \theta^\ast} \biggr|_{ \cos \theta^\ast = \tanh y^\ast }
     \right.
\nn
&
     \left.
      + f_{i}(\sqrt{\tau} e^Y )f_{j}( \sqrt{\tau} e^{-Y} )    
        \frac{ d \hat{\sigma}_{ij}}{d \cos \theta^\ast} \biggr|_{ \cos \theta^\ast = -\tanh y^\ast } 
     \right\} .
\end{align}
We have used the definitions
\begin{align}
Y  = \frac{y^{-} + y^{+}}{2} \hspace{1cm}  y^\ast = \frac{y^{-} - y^{+}}{2},
 \end{align}
and the allowed region is
\begin{align}
  - \log \frac{E_{\rm cm}^2}{M_{\mu \mu}^2}  &< y^{-} +  y^{+} < \log \frac{E_{\rm cm}^2}{M_{\mu \mu}^2} \nn
  - \infty  &< y^{-} -  y^{+} < \infty . 
\end{align}
The two terms in braces in Eq.~\ref{diff:xsec} represent the quark coming from either proton.  
The muon's rapidity can be replaced with its pseudorapidity ($y^\pm \to \eta^\pm$) at
the energies we are considering, 
Inserting Eq.~(\ref{diff:xsec}) into Eq.~(\ref{Eeta_def})
gives the average value of the ellipticity.   
We similarly calculate the center-edge asymmetry using
\begin{align}
\frac{d^2\sigma}{dM^2 d \cos \theta^{\ast}} = 
    \frac{1}{E_{\rm cm}^2}   
    \sum_{ij} 
    \frac{d\mathcal{L}_{ij}(\tau)}{d \tau}
        \frac{ d \hat{\sigma}_{ij}(\tau s) }{d \cos \theta^\ast} 
    \hspace{1cm}
    \tau = M^2/E_{\rm cm}^2, 
\end{align}
which is then inserted to Eq.~(\ref{Ace_def}).
Finally, when integrating over the invariant mass $M$, 
one chooses different integration regions depending on
the width of the resonance.
When the resonance is narrow, events are selected in a single invariant mass bin around the resonances,
which enhances signal over background.  
However, for broad resonances, there isn't a convenient mass window that enhances the sample,
and one is forced to include all of the events above a given invariant mass,
$M_{\mu \mu}^{\rm cut}$,
which will vary according to the shape and mass of the resonance.
We have chosen the values of $M_{\mu \mu}^{\rm cut}$ to be $400$ and $1000$ GeV
for the 7 and 10 TeV runs, respectively.


The appearance of (parity conserving) new physics above the mass cut will give larger values
for the ellipticity due to the larger proportion of of centrally peaked $\mu^-$s.  
For example, the representative RS model has a contribution to the DY cross section
from the initial state gluons.  
These gluons tend to have comparable momentum fractions and so there isn't a large boost as in 
the $q\bar{q}$ initial state; therefore, 
the $\eta^-$ distribution from $gg$ tends to be centrally peaked.  
The contribution from $q\bar{q} \to G^{\ast} \to \mu^+ \mu^-$ shows a slight asymmetry, but it is dwarfed
by the gluon contribution that washes out any distinctive shape.  

The rather high invariant mass cut means there are only 
100 or so events for 1 fb$^{-1}$ at 7 TeV and so the observables will 
have large statistical uncertainties.   
\begin{figure}[t]
\begin{center}
\includegraphics[scale=0.4]{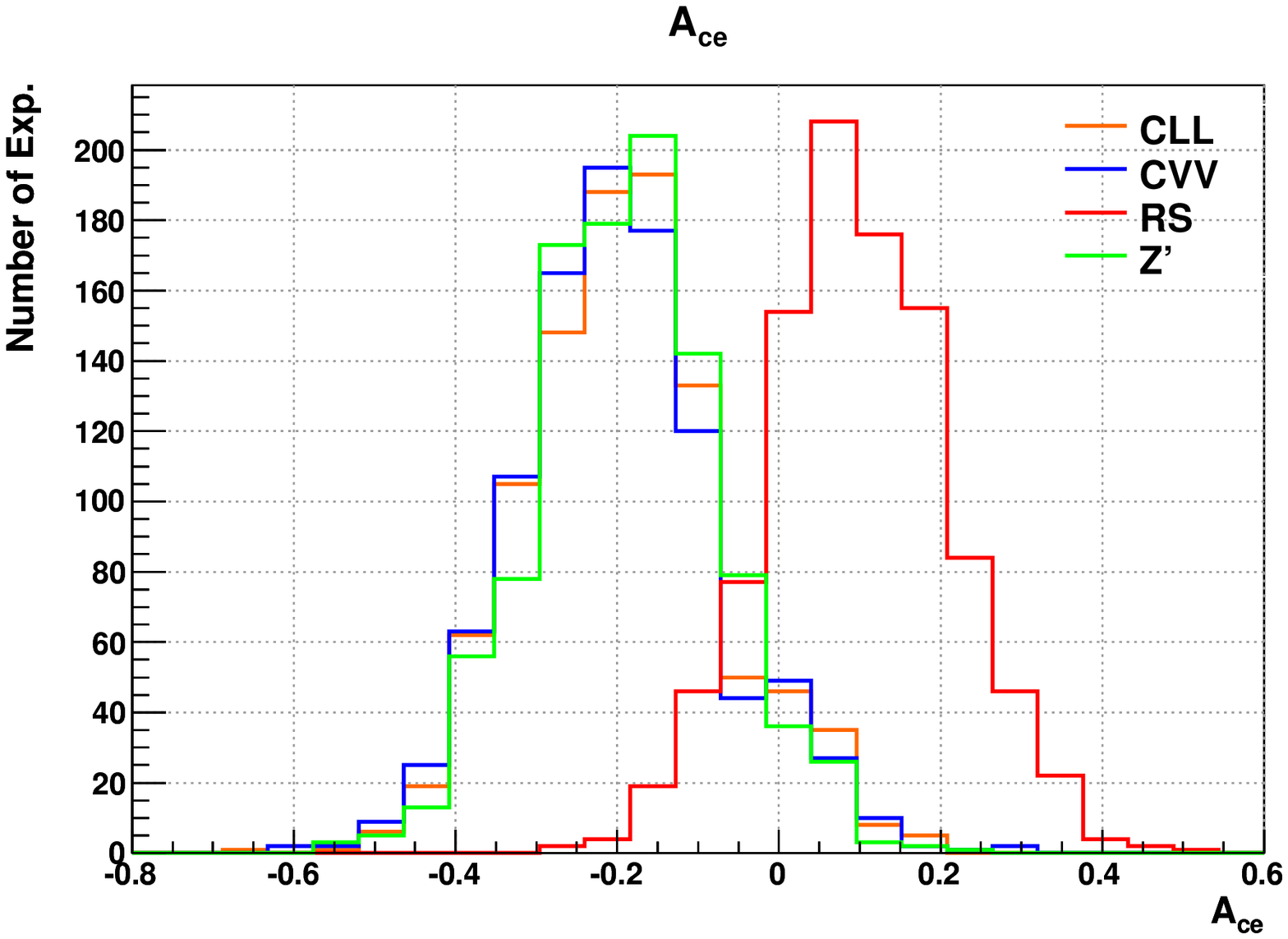}
\includegraphics[scale=0.4]{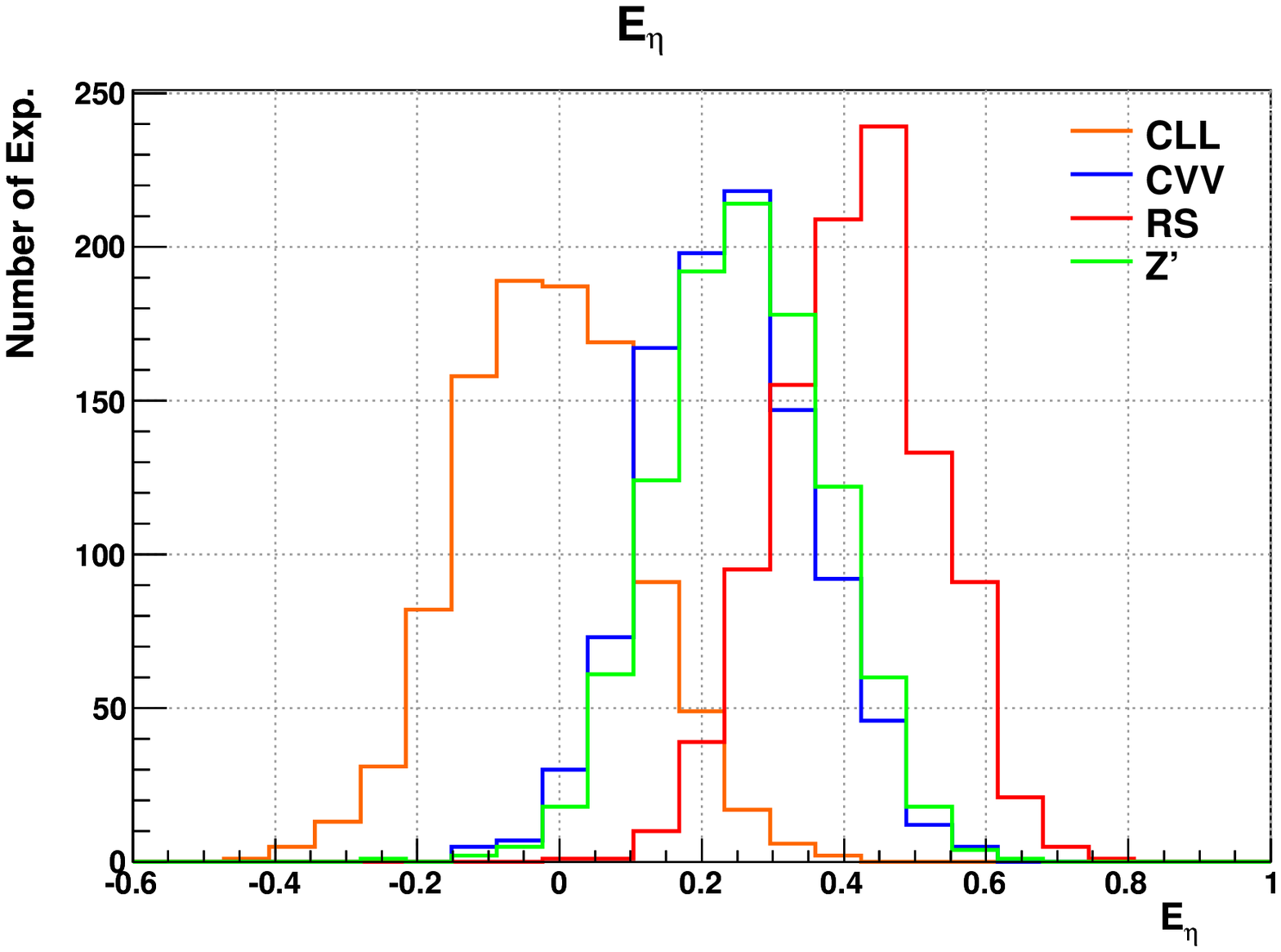}
\includegraphics[scale=0.4]{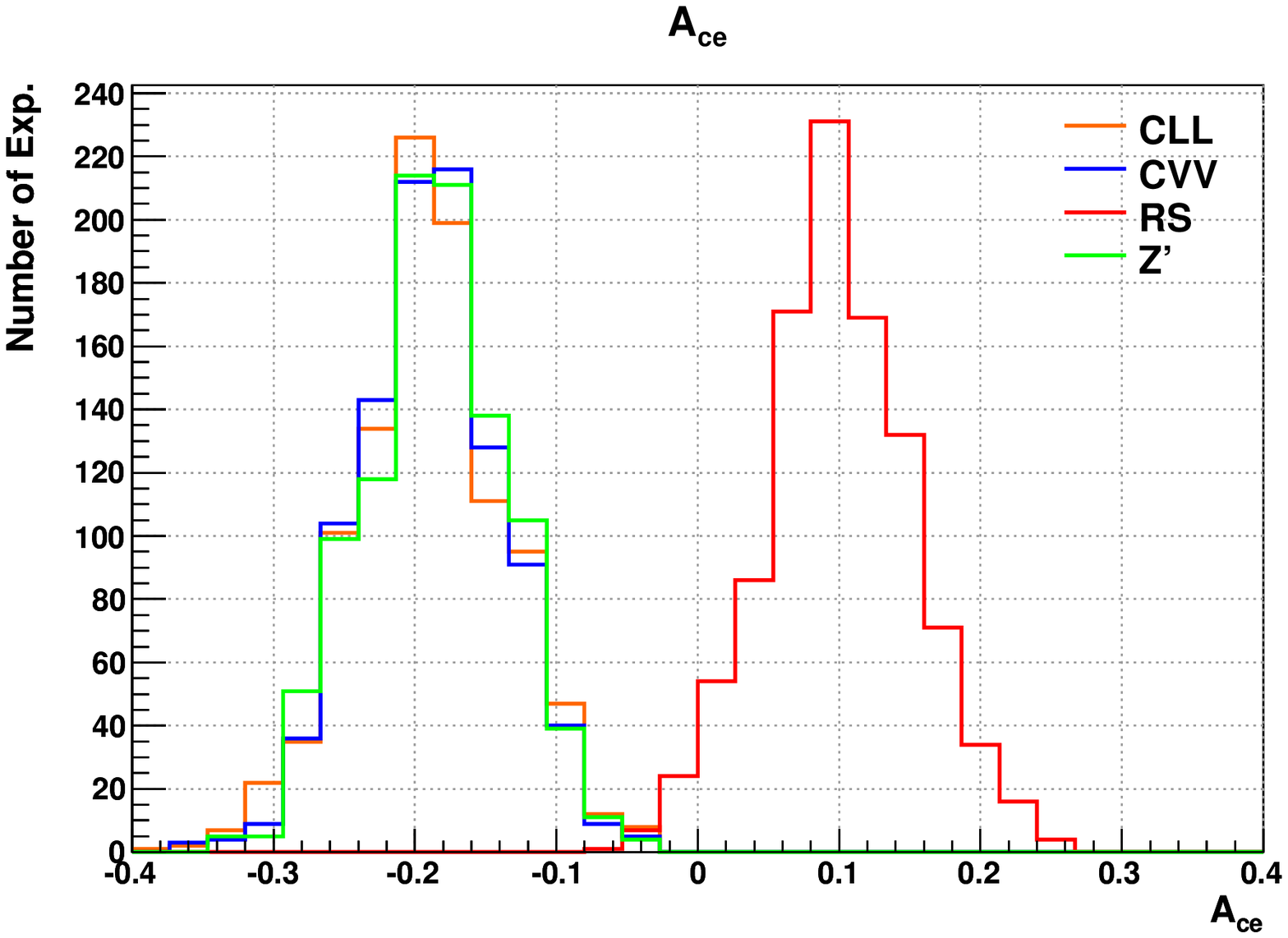}
\includegraphics[scale=0.4]{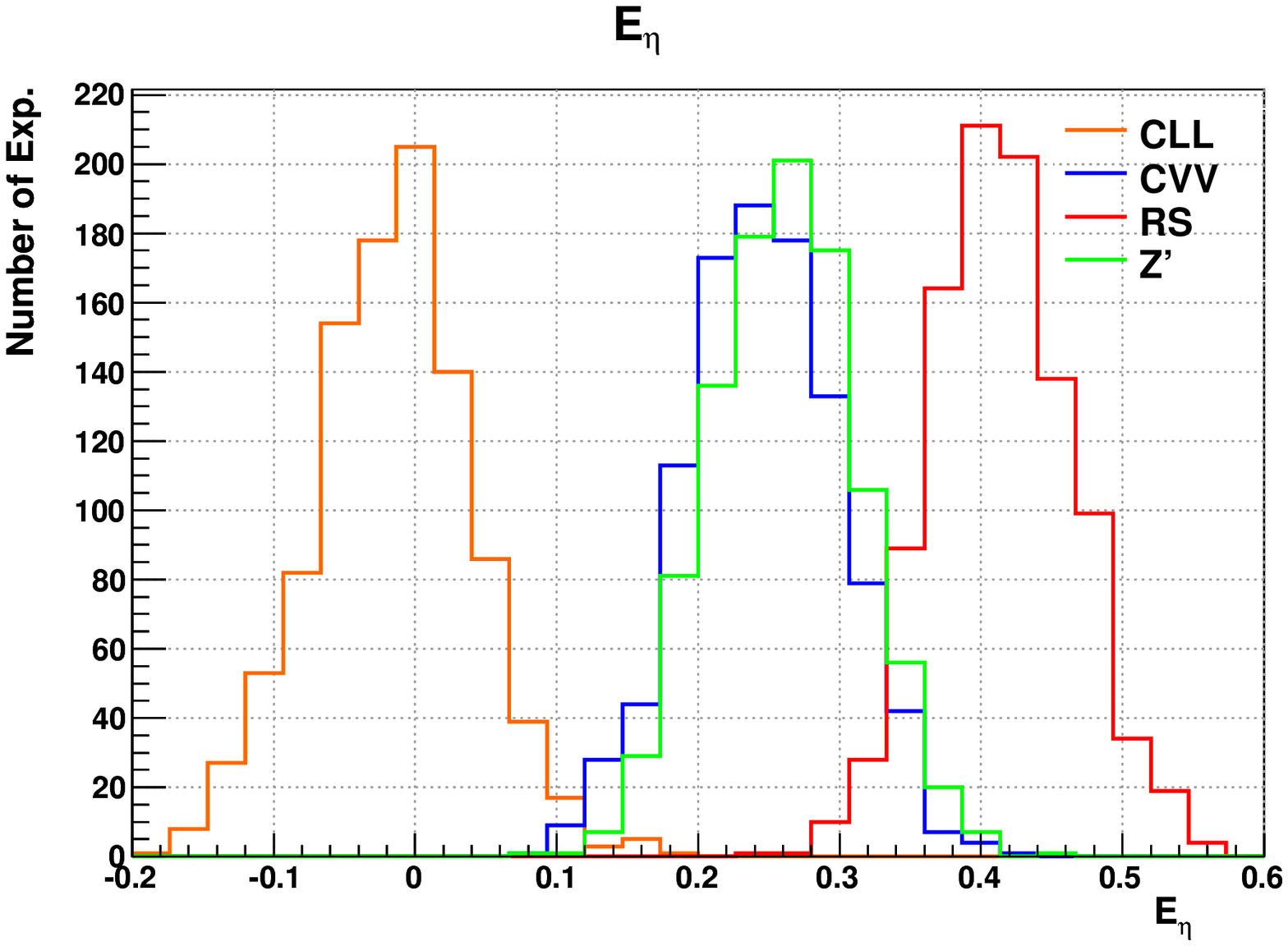}
\caption{$A_{\rm ce}$ (left) and $E_{\eta}$ are shown for 1000 pseudo-experiments for  
    $E_{\rm cm} = 10$ TeV, $\int dt \mathcal{L} = 1$ fb$^{-1}$ (upper) and 5 fb$^{-1}$ (lower) 
    with $M_{\mu \mu} > 1000$ GeV.
    RS model with $\Mg = 2000$ GeV, $k/\MPl = 0.5$ (red), 
    ($B-3L_\mu$) $Z'$ model with $\Mg = 2000$ GeV, $\e = 0.61$ (green), 
    LL composite model with destructive interference and $\Lambda = 5940$ GeV (blue),
    and
    VV composite model with destructive interference and $\Lambda = 8100$ GeV (orange), 
    }
\label{E:3}
\end{center}
\end{figure}
To model this uncertainty, a distribution of $A_{\rm ce}$ and $E_{\eta}$ was generated for 
1000 pseudo-experiments for 1, 5 and 10 fb$^{-1}$ at various CM
energies.  A representative set of distributions are shown in Fig.~\ref{E:1}.  
The pseudo-experiments were generated with Pythia 8  
\cite{Sjostrand:2006za, Sjostrand:2007gs}.
The LHC will only be able to make a single measurement and the distributions of Fig.~\ref{E:1} for the
completing models are too similar for effective discrimination.  There is some separation of the RS 
model from the rest, but to say the models are distinguishable is a bit of a stretch since there is
significant overlap with the other distributions.
We conclude that for masses greater than 1000 GeV, 
1 fb$^{-1}$ at 7 TeV is not enough integrated luminosity to be able to reliably distinguish
among the models using these observables.  Indeed, it is unlikely that any
angular distribution observable will be able to discriminate models with so few statistics. 

Discriminating power improves with more events,
whether due to higher cross section (due to lighter new physics or higher CM energy) or
higher integrated luminosity.
The distributions are presented in 
Fig.~\ref{E:3} for $E_{\rm cm} = 10$ TeV with 1 fb$^{-1}$ (upper plots) and 5 fb$^{-1}$ (lower plots)
of luminosity for the same models
as in Fig.~\ref{E:1}, but with modified couplings and higher resonance masses or composite scales.
The distributions in the lower plots show considerable separation indicating 
that for certain parameters, these models are distinguishable.  For example, if the experiment 
measured a value of $E_\eta$ of 0.5, there is little chance that the LL composite model
could be the underlying
physics. The $A_{\rm ce}$ distributions shows clear
separation between the RS model and the other spin 1 models, whereas the $E_{\eta}$ distribution
distinguishes between the RS model, the LL composite model, 
and the parity conserving $Z'$ and VV composite models.
This is a clear demonstration that $E_{\rm \eta}$ has more sensitivity to parity violation than $A_{\rm ce}$
and will be a useful observable for model discrimination for broad resonances. Finally, $A_{\rm ce}$ and $E_{\eta}$ are complementary variables.  Measuring $A_{\rm ce}$ will help discriminate between models of differing spin structure, after which a measurement of $E_{\eta}$ will
further differentiate models based on their parity violating character.
Once the models are distinguished, a more thorough investigation of the properties of the
resonance can be performed.

\section{Conclusions}

In gearing up to study new physics at the LHC, we must consider the different features of the new collider. 
We have shown that, even in the first LHC run, its energy advantage over the Tevatron will allow it to 
probe new resonance physics, but only when couplings are sufficiently large to give sufficiently many 
signal events at relatively low luminosity. This has led us to the search for broad resonances. We have 
seen that careful statistical analyses will let us not only search for new resonances, but
also discriminate them from contact interactions (as well as background).

We have furthermore considered angular distributions, 
a discriminator most likely to be useful only when
the LHC runs at higher energy and luminosity. 
We demonstrated that the ellipticity is sensitive to parity-violation,
even without knowing the quark direction, whereas the center-edge asymmetry
is only sensitive to the spin structure of the interaction.  Both observables
provide complementary, useful, information.

\section{Acknowledgements}
We would like to thank Matthew Schwartz, Emily Thompson and Bob Cahn for helpful discussions about the project.
We would also
like to thank Ryan Kelley for his expert help with Pythia 8 and ROOT.
LR is supported by NSF grant PHY-055611.

\appendix
\appendixpage

\section{Binning}\label{sec:binning}

We performed a binned analysis in section \ref{sec:results}, which introduces uncertainties
relative to an unbinned analysis that are proportional to the square of the bin size \cite{Cahn:Bin}.
 Nevertheless, we anticipate that the binned analysis gives comparable results to an
unbinned analysis for the following reasons:

\begin{figure}[!b]
\begin{center}
\includegraphics[scale=0.9]{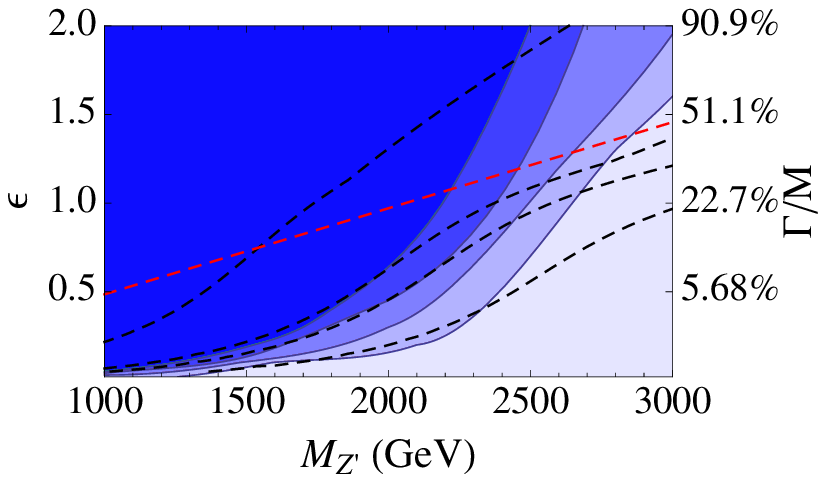}
\includegraphics[scale=0.9]{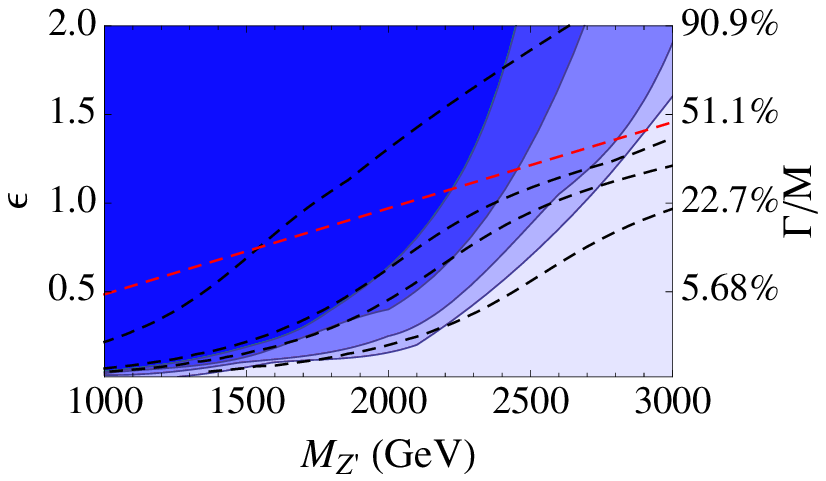}
\caption{Comparison of statistical analysis with different bin sizes for a $Z'$ compared
to a VV destructive contact interaction.  The reliability shadings and
cross sections are as in Fig.~\ref{fig:ZPstat}, and the analysis was performed with bins of 50 GeV 
(left) and 25 GeV (right) width.}
\label{fig:Bin}
\end{center}
\end{figure}

\begin{enumerate}

\item The features of a broad resonance are on a scale $\sim\Gamma$, which is must larger than
the 100 GeV bin width.

\item With few events at high invariant mass, we never expect to see more than a single event
(or two at most) per bin.

\item At low invariant mass, where we expect more events, the distribution is
dominated by SM interference, which we expect to be similar for different competing models of
new physics (such as as $Z'$ vs.~a destructively-interfering contact interaction.
A counting experiment suffices in this region, as we don't need high sensitivity
to the shape of the invariant mass distribution here.

\end{enumerate}

We present the results of a binned analysis with bin sizes 50 GeV and 25 GeV
in Fig.~\ref{fig:Bin}, which are reduced relative to our standard 100 GeV bin size.
It is evident that any differences between them are minor and do not affect our results
qualitatively; therefore, the 100 GeV bin size is sufficiently small
to give accurate results.

Reducing the bin size does have some advantages, however.  In particular,
the accuracy of the fit parameters is enhanced by performing an unbinned analysis, even
though $\Delta Q$ does not change much between them.
In our case, however, we are not actually concerned with a precise determination of the 
resonance parameters, but whether characteristics of the invariant mass distribution allow
us to generically distinguish it from other types of new physics, and this can be achieved accurately
using a binned analysis.

\section{Detector smearing}\label{sec:smearing}
The muon detectors at ATLAS and CMS have some finite resolution that introduces uncertainties
into the invariant mass measurements.  This has the tendency to smear the resulting distribution
that comes out of the detector.  The smearing should be applied both to the differential cross
section used to generate the simulated data, as well as the contact and resonance distributions
used to fit the data.  This makes the whole simulation/fitting procedure much more
computationally-intensive, particularly considering the number of fits that are required to scan
over the parameter space and generate the $\Delta Q$ distributions from the contact interactions.

\begin{figure}[t]
\begin{center}
\includegraphics[scale=0.9]{img/NUVVdestr}
\includegraphics[scale=0.9]{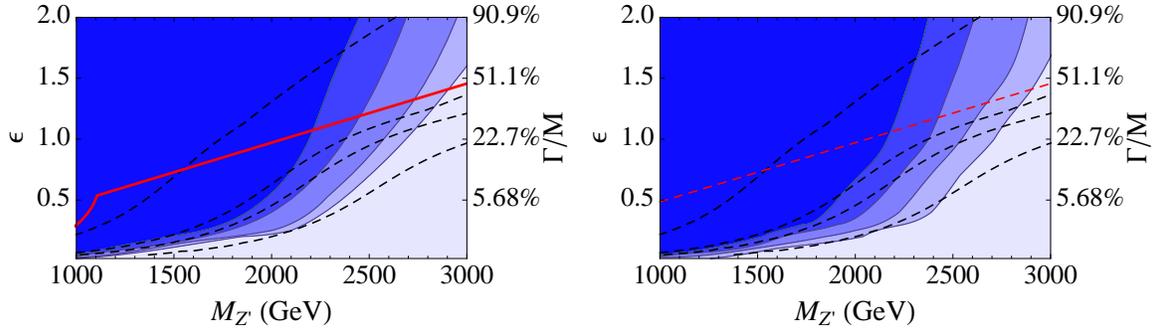}
\caption{Comparison of statistical analysis without (left) and with (right) Gaussian smearing of
the final muon states for the non-universal $Z'$ compared to a VV destructive contact
interaction.  The reliability shadings are as in Fig.~\ref{fig:ZPstat}, as are the
cross sections.}
\label{fig:Smear}
\end{center}
\end{figure}

Smearing tends to give
otherwise narrow resonances a width of about the detector resolution, which at
$\mathcal O(100\,\,\mathrm{GeV})$ is still too small to leave any doubt that the new physics
is a resonance.  For broad resonances, however, it could further obscure the existing
features of the resonance and make it look even more like a contact interaction.  To determine
the effects on our result, we repeated the analysis for a $Z'$ assuming Gaussian smearing
using the detector resolution given by \cite{Kortner:2007qj}.
Our result is shown in Fig.~\ref{fig:Smear}.

The introduction of detector smearing does change the results,
albeit by a relatively small margin (for broad resonances, we have $~50\,\,\mathrm{GeV}$
loss in discriminatory power at fixed $\epsilon$).
The detector effects do not, however, qualitatively change the results in any way,
introducing only a small, horizontal shift in the reliability curves.  Given that a proper
analysis at ATLAS or CMS will require full-scale modeling of the detector,
we are content to accept the non-smeared results as valid to within such a
$\sim 50\,\,\mathrm{GeV}$ window.

\end{document}